\documentclass[prl,reprint,twocolumn,floatfix,superscriptaddress]{revtex4-1}
\usepackage[utf8]{inputenc}
\usepackage{amsmath}
\usepackage{amssymb}
\usepackage{physics}                                % useful commands for typesetting QM
\usepackage{graphicx}
\usepackage{microtype}                              % minor typographical enhancements
\usepackage{siunitx}
\usepackage{color}
\usepackage[export]{adjustbox}
\usepackage[caption=false]{subfig}

\usepackage{hyperref}

\hypersetup{
    colorlinks = true,
    linkcolor  = blue,
    citecolor  = blue,
    urlcolor   = blue,
}

\newcommand{\mc}[1]{\mathcal{#1}}                   % shorthand
\renewcommand{\v}[1]{\mathbf{#1}}                   % shorthand

\begin{document}

\title{Steady-state superconductivity in electronic materials with repulsive interactions}
\author{O.~Hart}
\affiliation{T.C.M.~Group, Cavendish Laboratory,  J.J.~Thomson Avenue, Cambridge CB3 0HE, United Kingdom}
\author{G.~Goldstein}
\affiliation{T.C.M.~Group, Cavendish Laboratory,  J.J.~Thomson Avenue, Cambridge CB3 0HE, United Kingdom}
\affiliation{Physics and Astronomy Department, Rutgers University, Piscataway, NJ 08854, USA}
\author{C.~Chamon}
\affiliation{Physics Department, Boston University, Boston, Massachusetts 02215, USA}
\author{C.~Castelnovo}
\affiliation{T.C.M.~Group, Cavendish Laboratory,  J.J.~Thomson Avenue, Cambridge CB3 0HE, United Kingdom}
\date{October 2018}

\begin{abstract}
    We study the effect of laser driving on a minimal model for a hexagonal two-dimensional material with broken inversion symmetry.
    Through the application of circularly polarised light and coupling to a thermal free electron bath, the system is driven into a nonequilibrium steady state with asymmetric, nonthermal carrier populations in the two valleys.
    We show that, in this steady state, interband superconducting correlations between electrons can develop independent of the sign of the electron--electron interactions.
    We discuss how our results apply, for example, to transition metal dichalcogenides. This work opens the door to technological applications of superconductivity in a range of materials that were hitherto precluded from it.
\end{abstract}

\maketitle

%%%%%%%%%%%%%%%%%%%%%%%%%%%%%%%%%%%%%%%%%%%%%%%%%%%%%%%%%%%%%%%%%%%%%%
%%%%%%%%%%%%%%%%%%%%%%%%%%%% INTRODUCTION %%%%%%%%%%%%%%%%%%%%%%%%%%%%
%%%%%%%%%%%%%%%%%%%%%%%%%%%%%%%%%%%%%%%%%%%%%%%%%%%%%%%%%%%%%%%%%%%%%%

The breaking of inversion symmetry in two-dimensional materials can give rise to dramatic changes in their response to optical driving.
Such spatial symmetry breaking occurs naturally in the monolayer group-VI transition metal dichalcogenides (TMDs), which host two inequivalent but degenerate (due to time reversal symmetry) valleys at opposite edges of their hexagonal Brillouin zone (BZ)~\cite{Xiao2012}. It has been shown experimentally that the carrier populations in these two inequivalent valleys can be tailored individually using circularly polarised light~\cite{Zeng2012, Mak2012, Cao2012}, an effect known as circular dichroism. The robustness of the valley index in these materials, in part due to the large momentum separation of the valleys, has led to the rise of the new field of `valleytronics'~\cite{Schaibley2016}. We demonstrate how this valley-selective driving can also give rise to novel phase transitions. In particular, we show that interband superconducting correlations can develop for both attractive and repulsive interactions between electrons.

Nonequilibrium superconductivity has a long history~\cite{Langenberg1986}, beginning in the 1960s with the Wyatt-Dayem effect: experiments on thin films of aluminium and tin showed that irradiation with sub-gap microwaves gives rise to an increase in the superconducting gap, the critical current, and the critical temperature~\cite{Wyatt1966, Dayem1967}. These results were first explained by Eliashberg~\cite{Eliashberg1970}, who showed that these effects could be attributed to a redistribution of quasiparticles to higher energies in response to the microwave driving. Subsequent experiments showed that this mechanism could in fact lead to an enhancement of $T_c$ up to several times its equilibrium value~\cite{Pals1979, Blamire1991, Heslinga1993, Komissinski1996}.
In recent years, superconducting order has been shown to develop following femtosecond laser pulses in the cuprates~\cite{Fausti2011, Mankowsky2014, Hu2014, Kaiser2014} and other materials~\cite{Mitrano2016}. For an overview of the current state-of-the-art experiments and possible theoretical explanations, see Ref.~\cite{Cavalleri2018}.
\begin{figure}
    \centering
    \includegraphics[width=\linewidth]{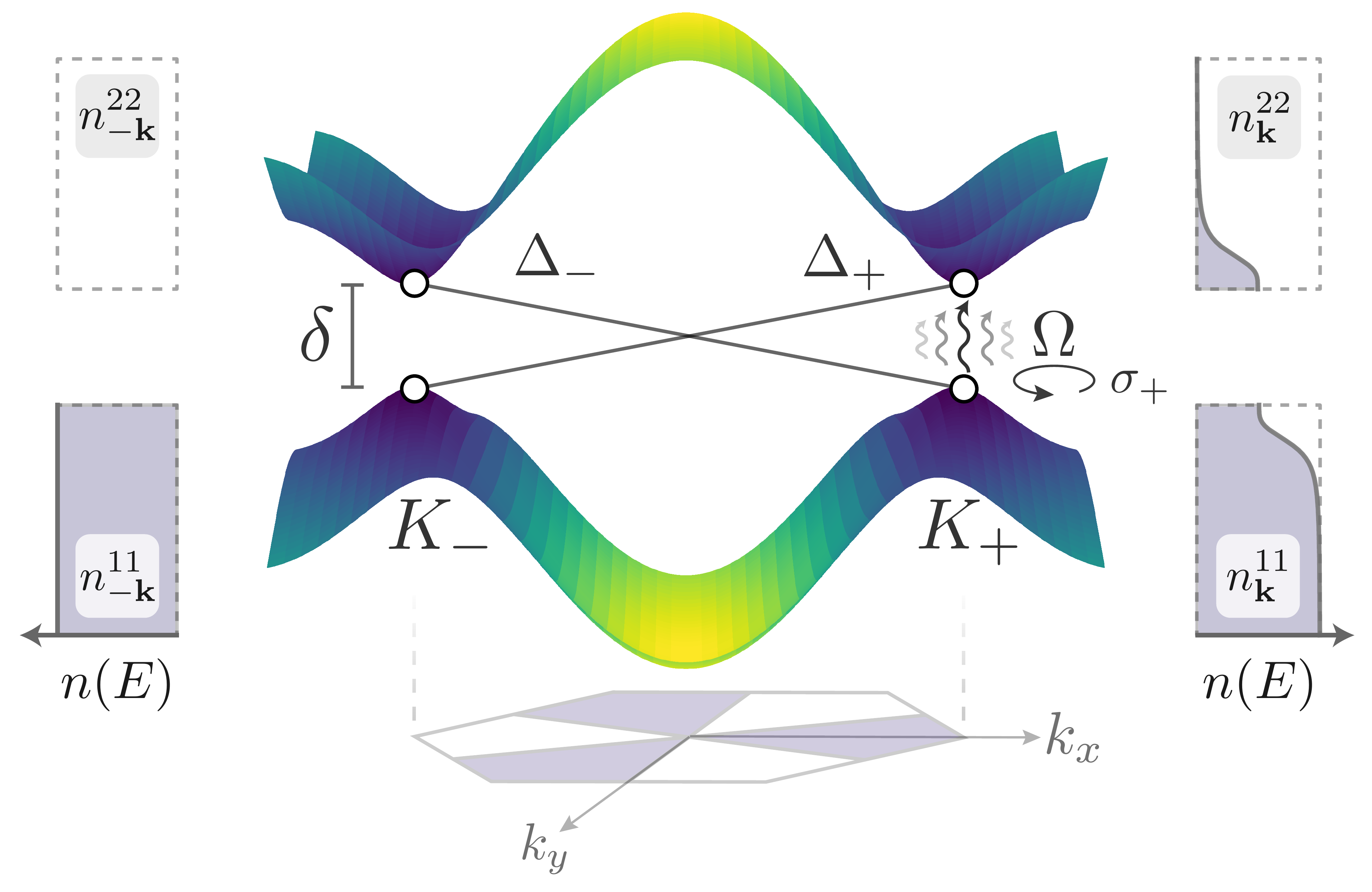}
    \caption{Schematic illustration of the pairing mechanism. The valley $K_+$ is driven with $\sigma_+$ polarised light of frequency $\omega_0 \simeq \delta$, the band gap, leading to a nonthermal population of the single-particle states near the centre of the valley. By virtue of broken inversion symmetry, valley $K_-$ is left unaffected by the laser.
    This induces a nontrivial population population difference between the upper and lower bands at $\pm \v{k}$. The corresponding occupations of the two valleys, $n(E)$, are illustrated qualitatively on their respective sides of the figure~\cite{pauliblockNote}. Our results show that one of the two pairing channels, $\Delta_+$ or $\Delta_-$, represented symbolically by the solid lines connecting the open circles, is always nonvanishing for sufficiently large $\Omega$.
    }
    \label{fig:schematic}
\end{figure}

Within standard BCS theory~\cite{BCS}, the self-consistency condition determining the superconducting gap $\Delta$ in a single-band superconductor may be written as
\begin{equation}
    1 = -\frac{V}{N} \sum_{\v{k}} \frac{1-n_\uparrow(E_\v{k}) - n_\downarrow(E_\v{k})}{2E_\v{k}}
    \, ,
    \label{eqn:bcs-self-consistency}
\end{equation}
where $V$ characterises the strength (and sign) of the electron--electron interaction, $E_\v{k} = \sqrt{\Delta^2 + \xi_\v{k}^2}$ is the quasiparticle energy, and $n_\sigma(E_\v{k})$ is the occupation of the quasiparticle state $\sigma$ at energy $E_\v{k}$. In thermal equilibrium at temperature $T$, the occupation numbers satisfy $1-n_\uparrow-n_\downarrow = \tanh(E_\v{k} / 2T)$, requiring an attractive interaction between electrons for a superconducting instability to develop.

In this letter we consider a two-band ($\alpha = 1, 2$) system subject to interband pairing interactions (see Fig.~\ref{fig:schematic}), which may be described by~\eqref{eqn:bcs-self-consistency} with the replacements $n_\uparrow \to n_{\v{k}}^{22}$ and $n_\downarrow \to n_{-\v{k}}^{11}$, i.e., the occupation numbers of the two bands.
In equilibrium, the lower band is fully populated, $n_{\v{k}}^{11} = 1$, and the upper band is empty, $n_{\v{k}}^{22} = 0$, implying that the population difference $1-n_{\v{k}}^{22}-n_{-\v{k}}^{11}$ appearing in~\eqref{eqn:bcs-self-consistency} vanishes.
Through valley-selective driving in $\v{k}$-space, one can populate the upper band, $n_{\v{k}}^{22} > 0$, over some region of momentum space (in our system, one of two valleys) at the expense of depleting the lower band, $n_{\v{k}}^{11}<1$.
The essential idea of this work is that the population differences for the two valley--band channels depicted in Fig.~\ref{fig:schematic} have opposite signs, namely $1-n_{\v{k}}^{22}-n_{-\v{k}}^{11} < 0$ and $1-n_{-\v{k}}^{22}-n_{\v{k}}^{11} > 0$.
One channel will then \emph{always} exhibit a superconducting instability, irrespective of the sign of $V$.

A similar mechanism was explored in the context of three-dimensional direct
band gap semiconductors in Ref.~\cite{Goldstein2015}. It was found
that interband superconductivity can be induced in such systems by
driving them away from equilibrium, even in the presence of repulsive
interactions between electrons. However, the occurrence of
superconductivity with repulsive interactions requires that certain
conditions be concomitantly satisfied: a resonance condition where
valence and conduction bands have opposite velocities over a range of
momenta; and a particular sign for the product of the difference in
curvature between the two bands at the resonance and the difference in
the escape times of the excited particles in the two
bands into the bath.

It is the goal of this letter to present a modified mechanism
for interband superconductivity which leads to nonzero superconducting
correlations without such restrictive requirements,
especially those associated with the system--bath parameters. The
mechanism that we present below for hexagonal two-dimensional
materials with broken inversion symmetry is robust in that there is
always one out of two channels that will lead to superconductivity with
repulsive interactions.

%%%%%%%%%%%%%%%%%%%%%%%%%%%%%%%%%%%%%%%%%%%%%%%%%%%%%%%%%%%%%%%%%%%%%%
%%%%%%%%%%%%%%%%%%%%%%%%%%%% CALCULATION %%%%%%%%%%%%%%%%%%%%%%%%%%%%%
%%%%%%%%%%%%%%%%%%%%%%%%%%%%%%%%%%%%%%%%%%%%%%%%%%%%%%%%%%%%%%%%%%%%%%

\textit{Model.}---For simplicity, we focus on a nearest-neighbour tight-binding model on a hexagonal lattice with the Hamiltonian
\begin{equation}
    H(\v{k}) = 
    \begin{pmatrix}
        \delta/2 & h(\v{k}) \\
        h^*(\v{k}) & -\delta/2
    \end{pmatrix}
    \, ,
    \label{eqn:tight-binding-model}
\end{equation}
where $h(\v{k}) = -t\sum_{i} e^{i\v{k}\cdot \v{d}_i}$, the vectors $\v{d}_{1,2}=\tfrac{a}{2}\v{y} \pm \tfrac{\sqrt{3}a}{2}\v{x}$, $\v{d}_3 = -a\v{y}$ connect nearest neighbours~\cite{Castro2009}, and $\delta > 0$ represents a staggered chemical potential. We henceforth set the distance between neighbouring atoms $a=1$.
The band structure $E_{\v{k}\alpha}$ corresponding to \eqref{eqn:tight-binding-model} has two bands ($\alpha = 1,2$, valence and conduction) separated by a gap $\delta$. The familiar Dirac cones of graphene, centred at $\v{K}_\pm = \pm \tfrac{4\pi}{3\sqrt{3}}\v{x}$, become gapped valleys in the presence of the staggered chemical potential. At the Dirac points $K_\pm$, there is an exact selection rule for optical band-edge transitions: circularly polarised light with polarisation $\sigma_\pm$ couples only to transitions within the $K_\pm$ valley~\cite{Yao2008}. Hence, each valley can be driven independently.
\begin{figure}
    \centering
    \subfloat[$\delta/t=1/4$]{\includegraphics[width=0.3\linewidth]{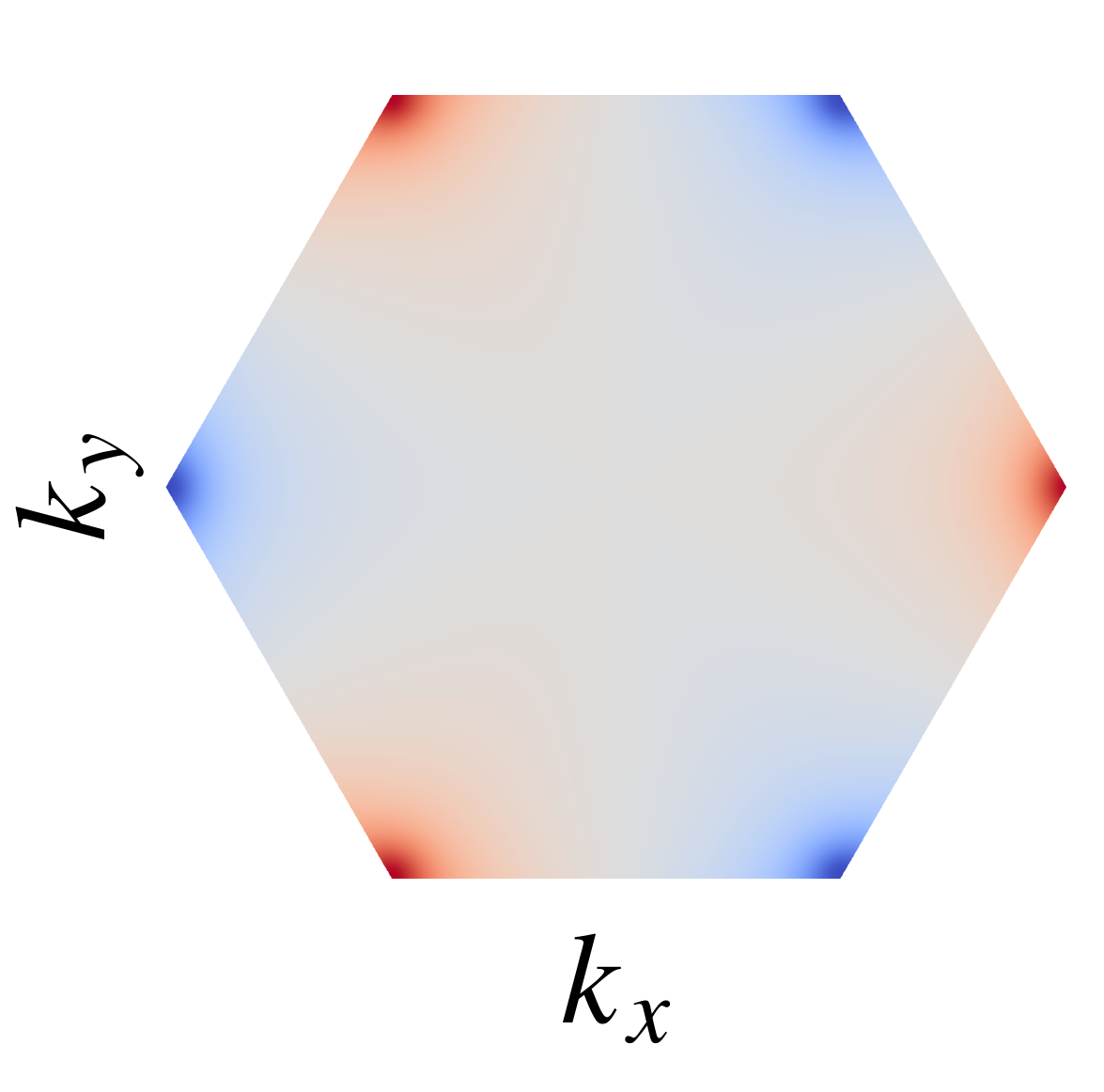}}
    \hfill
    \subfloat[$\delta/t=1$]{\includegraphics[width=0.3\linewidth]{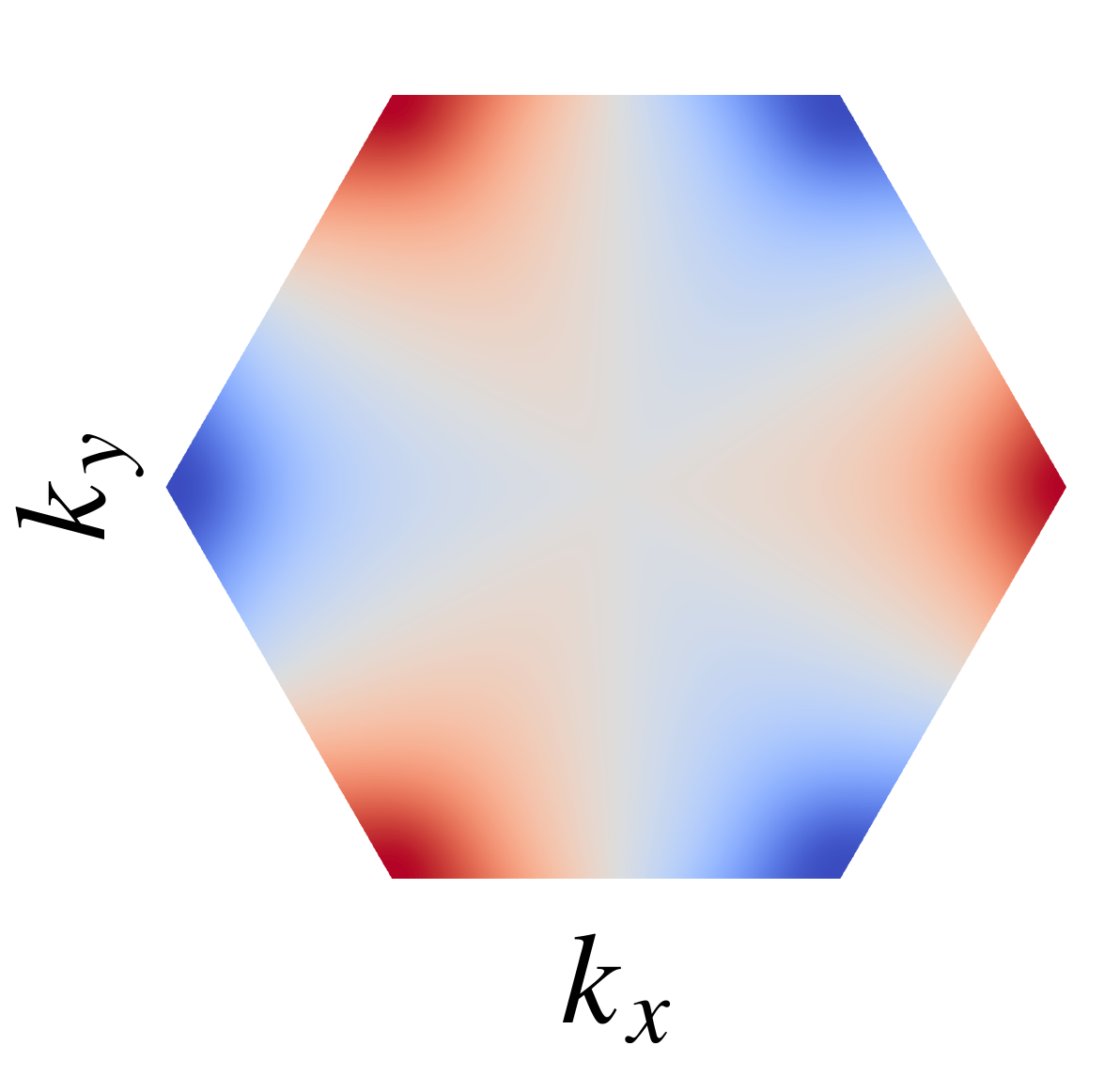}}
    \hfill
    \subfloat[$\delta/t=5$\label{fig:driving-pattern:large-gap}]{\includegraphics[width=0.4\linewidth]{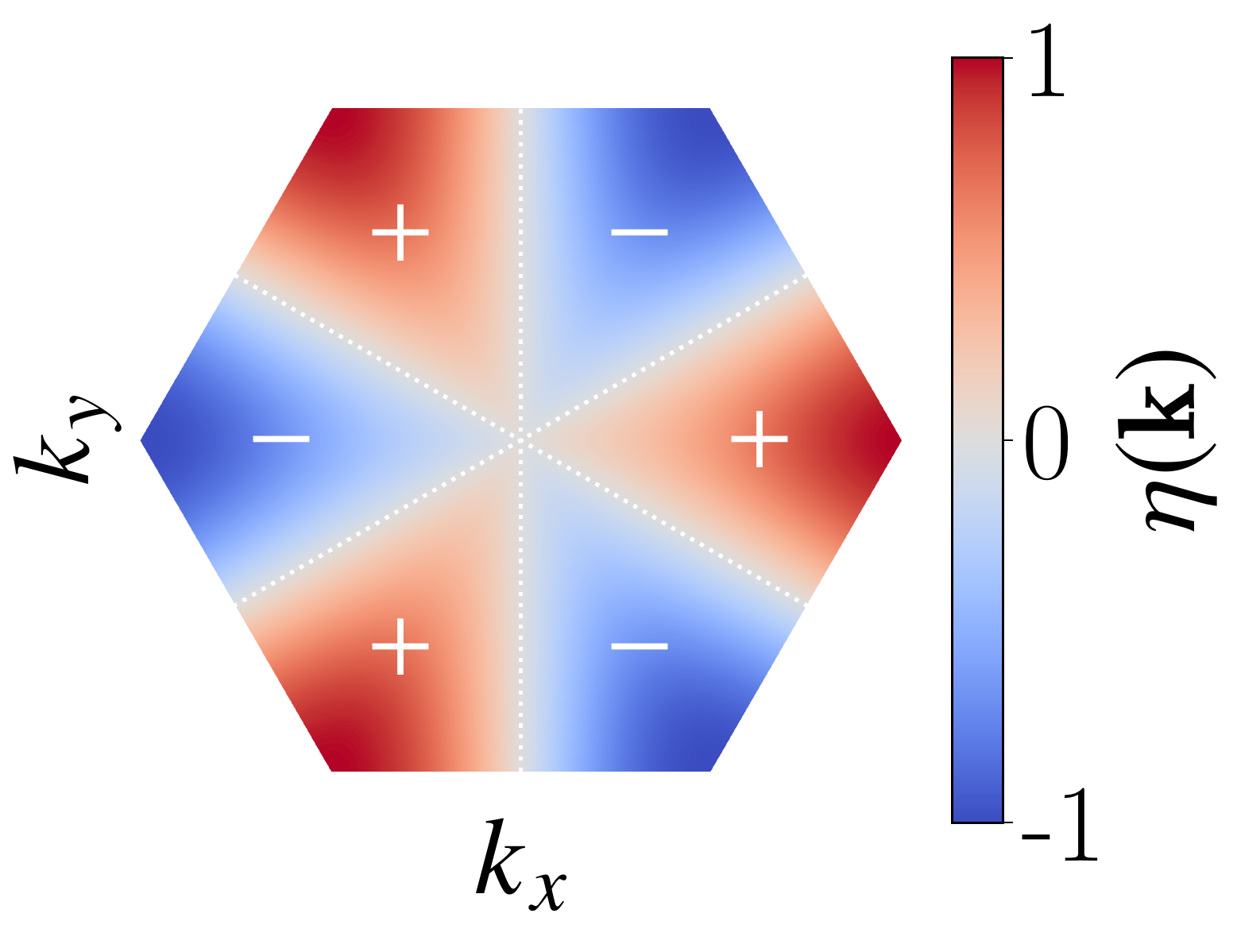}}
    \caption{The asymmetry, quantified by $\eta(\v{k})$, between absorption of light with circular polarisation $\sigma_+$ ($\eta = 1$) and $\sigma_-$ ($\eta = -1$). The valleys $K_\pm$, centered on $\v{K}_\pm = \pm \tfrac{4\pi}{3\sqrt{3}}\v{x}$, couple only to $\sigma_\pm$ polarisations, respectively. The plot is calculated for hexagonal materials described by the Hamiltonian \eqref{eqn:tight-binding-model} (see Supplementary Material).}
    \label{fig:driving-pattern}
\end{figure}

This asymmetry between absorption of $\sigma_\pm$ polarised light is quantified by the degree of circular polarisation $\eta(\v{k})$~\cite{Yao2008, Cao2012},
\begin{equation}
    \eta(\v{k}) = \frac{\abs*{\mc{P}_+^{21}(\v{k})}^2-\abs*{\mc{P}_-^{21}(\v{k})}^2}{\abs*{\mc{P}_+^{21}(\v{k})}^2+\abs*{\mc{P}_-^{21}(\v{k})}^2}
    \, ,
\end{equation}
where $\mc{P}^{21}_\pm(\v{k}) = \mel*{\psi_{2\v{k}}}{p_\pm}{\psi_{1\v{k}}}$ and $p_\pm = p_x \pm i p_y$ describe optical transitions between the conduction and valence bands. The asymmetry, calculated using~\eqref{eqn:tight-binding-model}, is plotted for various staggered chemical potentials over the first BZ in Fig.~\ref{fig:driving-pattern}. The selection rule is always exact ($\eta = \pm 1$) at $\v{K}_\pm$ for any nonzero $\delta$~\cite{Yao2008}, and for $\delta \gtrsim t$ the asymmetry spreads towards the centre of the BZ. The driving strength is parameterised in terms of the Rabi frequency $\Omega_\v{k} = (eE_0/2 m\omega_0)\mc{P}_\pm^{21}(\v{k})$, which we take to be real. $E_0$, $e$ and $m$ describe the strength of the electric field and the electronic charge and mass, respectively.

We will study two limiting cases: (i) when relaxation occurs
exclusively through tunnel coupling to a three-dimensional substrate,
and (ii) when fast intraband relaxation establishes a local
equilibrium in the upper and lower bands separately. The latter case is 
important for its closer connection to experiment, but the derivation of the 
results uses a more phenomenological approach that is easier to follow after 
exposure to the results of the former. Hence, for clarity of presentation, 
we will focus the discussion mainly on the former case where we are able to 
confirm our results using two separate methods, and present the derivation 
of the latter in the Supplementary Material. 

We assume a simplified driving pattern as a minimal model of $\sigma_+$ polarised driving in which $\Omega_\v{k} = \Omega$ in the regions of the first BZ where $\eta(\v{k})>0$ in Fig.~\ref{fig:driving-pattern:large-gap}, and $\Omega_\v{k}=0$ in the regions where $\eta(\v{k}) < 0$.
These two regions will be referred to as $\v{k} \in K_\pm$, respectively.
Although the Rabi frequency will in any real material depend continuously on momentum, in practice this dependence may be neglected since the dominant contribution to the superconducting gap equation comes from the vicinity of the surface $S_{\omega_0}=\{ \v{k} : E_{\v{k}2} - E_{\v{k}1} = \omega_0 \}$ where the laser is resonant.

Our complete model Hamiltonian is composed of an interacting system (S), a bath (B), and a system--bath (S--B) interaction
\begin{equation}
    H = H_\text{S} + H_\text{int} + H_\text{S--B} + H_\text{B}
    \, ,
\end{equation}
where
\begin{align}
    H_\text{S} &= \sum_{\lambda} E_{\lambda} c_{\lambda}^\dagger c_{\lambda}^{\phantom{\dagger}} + \sum_\v{k} \Omega_\v{k}  ( e^{i\omega_0t} c_{\v{k}2}^\dagger c_{\v{k}1}^{\phantom{\dagger}} + e^{-i\omega_0 t} c_{\v{k}1}^\dagger c_{\v{k}2}^{\phantom{\dagger}} ) \, , \\
    H_\text{int} &= \frac{1}{N} \sum_{\v{k}, \v{k}^\prime} V_{\v{k}\v{k}^\prime} c_{\v{k}2}^\dagger c_{-\v{k}1}^\dagger c_{-\v{k}^\prime 1}^{\phantom{\dagger}} c_{\v{k}^\prime 2}^{\phantom{\dagger}} \, , \label{eqn:interaction-hamiltonian} \\
    H_\text{S--B} &= \sum_{\lambda, n} t_{\lambda} (c_{\lambda}^\dagger a_{\lambda n}^{\phantom{\dagger}} + a_{\lambda n}^\dagger c_{\lambda}^{\phantom{\dagger}}) \, , \label{eqn:system-bath-interaction} \\
    H_\text{B} &= \sum_{\lambda, n} \omega_{\lambda n} a_{\lambda n}^\dagger a_{\lambda n}^{\phantom{\dagger}} \, .
\end{align}
The index $\lambda = (\v{k}, \alpha)$ labels the noninteracting system modes, and $N$ is the number of unit cells. Both the system and the bath are composed of spinless fermionic degrees of freedom: $\acomm*{c^{\phantom{\dagger}}_\lambda}{c^\dagger_{\lambda^\prime}}=\delta_{\lambda\lambda^\prime}$, and $\acomm*{a^{\phantom{\dagger}}_{\lambda n}}{a^\dagger_{\lambda^\prime m}}=\delta_{\lambda\lambda^\prime}\delta_{nm}$. 
(The assumption of spinlessness is made for simplicity but can be relaxed without changing our results---see Supplementary Material.) 
The system is driven by a laser of frequency $\omega_0$ (included semiclassically), and interacts via the scattering of interband pairs. Coupling the system to a bath with which it can exchange both energy and particles brings our system towards a unique nonequilibrium steady state.

%%%%%%%%%%%%%%%%%%%%%%%%%%%%%%%%%%%%%%%%%%%%%%%%%%%%%%%%%%%%%%%%%%%%%%
%%%%%%%%%%%%%%%%%%%%%%%%%%%% BORN MARKOV %%%%%%%%%%%%%%%%%%%%%%%%%%%%%
%%%%%%%%%%%%%%%%%%%%%%%%%%%%%%%%%%%%%%%%%%%%%%%%%%%%%%%%%%%%%%%%%%%%%%

\textit{Born--Markov approximation.}---The simplest possible analysis of our time-dependent Hamiltonian can be performed by moving into the frame rotating at $\omega_0$ and applying the Born--Markov approximation. In this approach, we assume that the baths have a continuous density of states $\nu_\lambda(\epsilon)$, and that they interact weakly with the system: $\pi{|t_\lambda|}^2 \ll \delta$. The dynamics of the system S, described by its reduced density matrix $\rho_\text{S}=\Tr_\text{B}\rho$, is then determined approximately~\footnote{This expression for the time evolution of $\rho(t)$ is not strictly correct, as it does not account for any changes in the quasiparticles of the system. The full Keldysh description discussed later in the text and in the Supplementary Material fixes this shortcoming.} by the Master equation~\cite{Puri2001}
\begin{multline}
    \dv{\rho_\text{S}}{t} = -i [H_\text{S}, \rho_\text{S}] + \sum_{\lambda} \Gamma_\lambda \big\{ n_\text{F}(\xi_\lambda) \mc{D}[c_\lambda^\dagger]\rho_\text{S} \\ + [1-n_\text{F}(\xi_\lambda)] \mc{D}[c_\lambda] \rho_\text{S} \big\} \, ,
    \label{eqn:born-markov-time-evolution}
\end{multline}
where $n_\text{F}(\xi) = (1+e^{\beta\xi})^{-1}$ is the Fermi--Dirac distribution, $\xi_\lambda = E_\lambda - \mu$, and the rates $\Gamma_{\lambda}=2\pi {|t_{\lambda}|}^2 \nu_\lambda(\xi_\lambda)$ are given by Fermi's golden rule.
The Lindbladian dissipators $\mc{D}$ are defined as $\mc{D}[X]\rho = (2 X\rho X^\dagger - X^\dagger X\rho - \rho X^\dagger X )/2$. We have neglected any Lamb shift corrections to \eqref{eqn:born-markov-time-evolution} which renormalise the band structure $E_\lambda$~\cite{Schlosshauer2007}. We will henceforth assume that the upper and lower bands are characterised by momentum-independent rates $\Gamma_\lambda \to \Gamma_\alpha$, $\alpha=1, 2$.

After making a mean field approximation for the superconducting order parameter in \eqref{eqn:interaction-hamiltonian}, we can write down the equations of motion for the populations and correlators, $n_\v{k}^{\alpha\beta}(t) = \ev*{c^\dagger_{\v{k}\alpha} c^{\phantom{\dagger}}_{\v{k}\beta}}$ and $s_\v{k}^{\alpha\beta}(t) = \ev*{c_{\v{k}\alpha}^\dagger c_{-\v{k}\beta}^\dagger}$, and solve for the steady state in the long-time limit (relaxation to this steady state typically occurs dynamically over timescales set by $\Gamma^{-1}_1$ and $\Gamma_2^{-1}$). One may then substitute the steady-state value for the anomalous correlator $s_\v{k}^{21}$ into the self-consistency condition
\begin{equation}
    \Delta_\v{k} = \frac{1}{N} \sum_{\v{k}^\prime} V_{\v{k} \v{k}^\prime} \expval*{c_{-\v{k}^\prime 1} c_{\v{k}^\prime 2}}
    \, ,
\end{equation}
for the order parameter. We make the following simplifying assumption about the scattering amplitudes $V_{\v{k}\v{k}^\prime}$: there exist only two relevant average scattering amplitudes $V$ and $V^\prime=ve^{i\phi}$ which, respectively, correspond to intra- ($K_\pm \to K_\pm$) and inter-valley ($K_\mp \to K_\pm$) scattering events. This in turn implies that there are only two momentum components of the gap, $\Delta_\pm$, corresponding to momenta in the vicinity of valley $K_\pm$. These two amplitudes will satisfy ${|V|} \gg {|V^\prime|}$; since the two valleys are separated by a large momentum transfer, intervalley scattering events are strongly suppressed [$V_{\v{K}_+,\v{K}_-}=0$ identically using the eigenstates of $H(\v{k})$ in~\eqref{eqn:tight-binding-model}] with respect to intravalley events. Using the Born--Markov equations of motion derived from \eqref{eqn:born-markov-time-evolution}, we obtain that
\begin{equation}
    \begin{aligned}
        \bar{\Delta}_\pm =  & -\bar{\Delta}_\pm \frac{V}{N}  \sum_{\v{k}\in K_\pm}   \frac{E_\v{k}}{E_\v{k}^2 + \Gamma^2} (1-n^{22}_{\v{k}} - n^{11}_{-\v{k}}) \\
          & -\bar{\Delta}_\mp  \frac{ve^{\pm i\phi}}{N}  \sum_{\v{k}\in K_\mp}   \frac{E_\v{k}}{E_\v{k}^2 + \Gamma^2} (1-n^{22}_{\v{k}} - n^{11}_{-\v{k}})
        \, ,
        \label{eqn:self-consistency-plus}
    \end{aligned}
\end{equation}
which is to be contrasted with the standard self-consistency condition \eqref{eqn:bcs-self-consistency}; the equilibrium populations have been replaced by their nonequilibrium counterparts. We have defined $E_\v{k} = \xi_{\v{k}1} + \xi_{\v{k}2}$, $\epsilon_\v{k} = \xi_{\v{k}2} - \xi_{\v{k}1} - \omega_0$ and $\Gamma = \Gamma_1 + \Gamma_2$. Note that \eqref{eqn:self-consistency-plus} reduces to the standard self-consistency condition when $\Gamma \to 0^+$, as it must.

\begin{figure*}
    \hspace*{\fill}
    \subfloat[\label{fig:critical-coupling-a}]{\includegraphics[width=0.475\linewidth,valign=c]{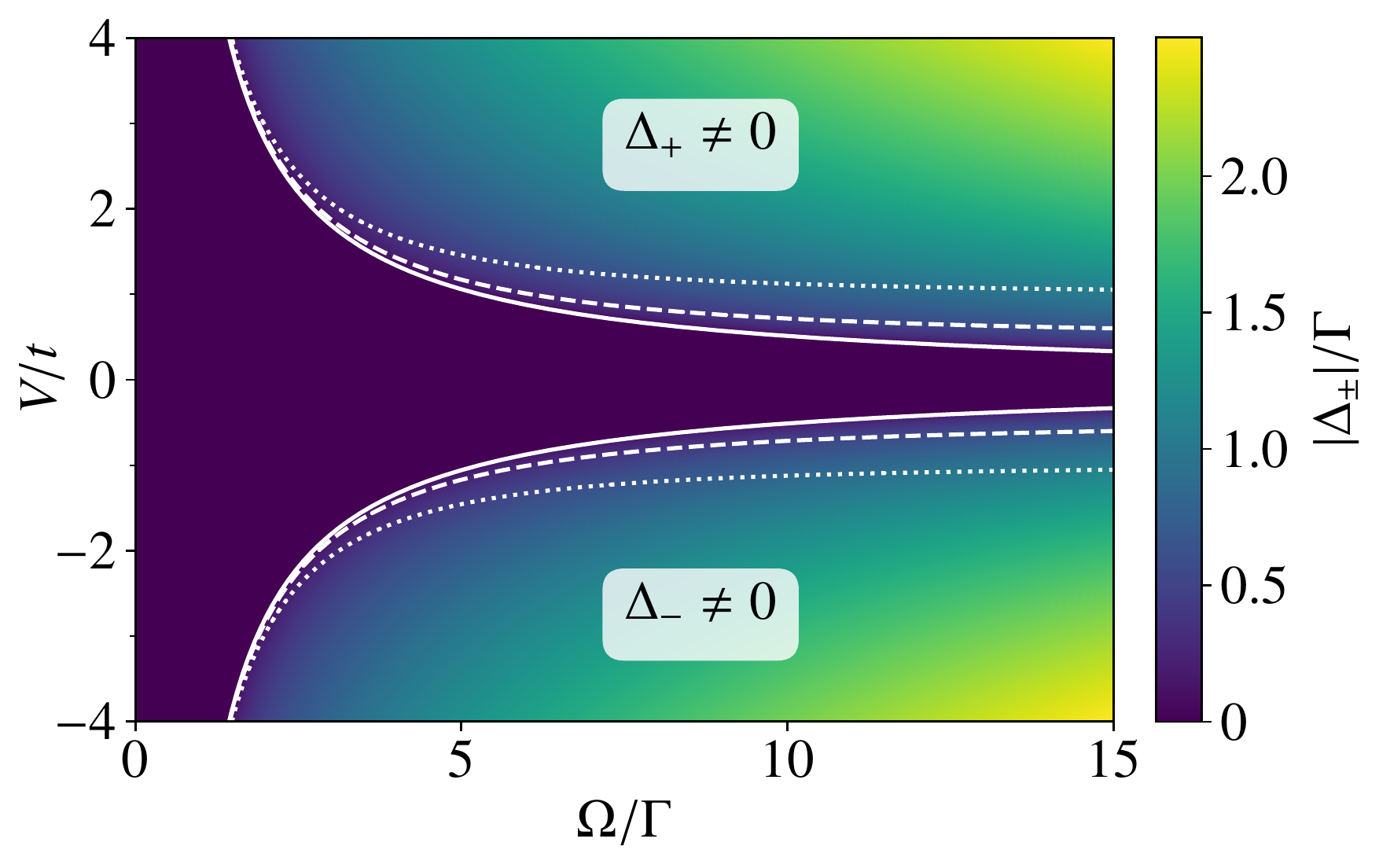}}%
    \hfill
    \subfloat[\label{fig:critical-coupling-b}]{\includegraphics[width=0.475\linewidth,valign=c]{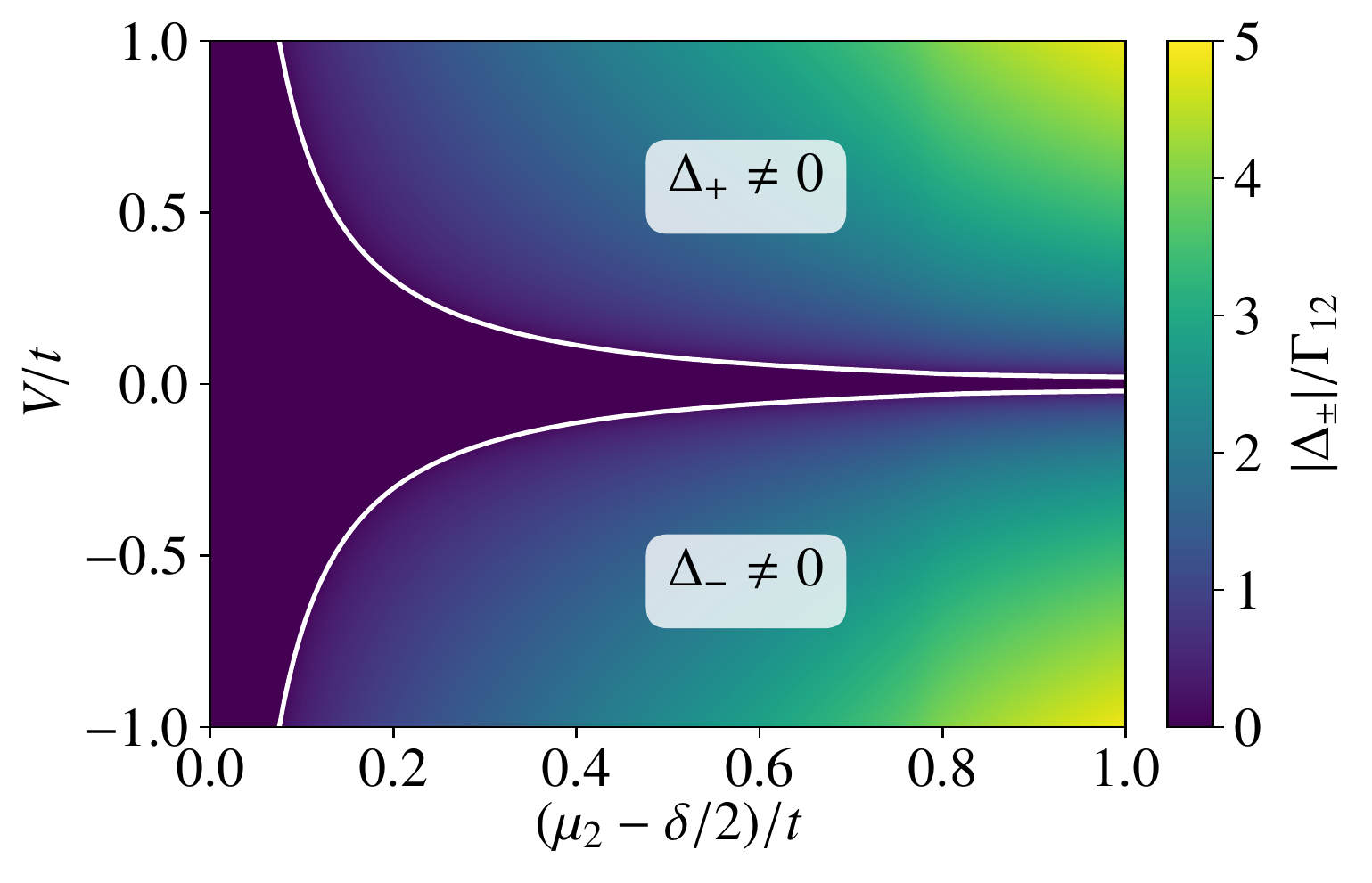}}%
    \hspace*{\fill}
    \caption{(a) Critical coupling $V_c$, in units of the hopping integral $t$, as a function of driving strength, parameterised by the Rabi frequency $\Omega$. There are two branches; one positive and one negative, which means that superconductivity may develop irrespective of the sign of interactions $V$. For sufficiently large driving (with respect to the damping $\Gamma$), the critical coupling saturates to $|V^\prime|$, the intervalley scattering matrix element. $V^\prime/t = 0, 1/2, 1$ correspond to the solid, dashed and dotted lines, respectively. If $V^\prime = 0$ (the value used for the colourmap) then only one of $\Delta_+$ or $\Delta_-$ is nonzero. A band gap of $\delta/t=5$, damping rates $\Gamma_1 = \Gamma_2 = 10^{-3}t$ and chemical potential $\mu=-\Gamma/2$ were used for the plot. (b) The equivalent plot for the case of fast intraband relaxation with rate $\Gamma_{12}$. The population difference $1-n_{\v{k}}^{22} -n_{-\v{k}}^{11}$ is now controlled by $\mu_2$, an effective chemical potential which determines the nonequilbrium populations of the $K_+$ valley. The parameters used for the plot are $\delta/t = 1/4$, $\Gamma_{12}=10^{-3}t$ and $\mu=-\Gamma_{12}/2$. With these parameters, an effective chemical potential of $\mu_2/t \simeq 0.2$ corresponds to $2.6\%$ polarisation of the $K_+$ valley.}
    \label{fig:critical-coupling}
\end{figure*}

In writing down \eqref{eqn:self-consistency-plus}, we have made the assumption that the damping $\Gamma$ is small. If $\Gamma$ is increased in magnitude, the gap parameters acquire a nontrivial oscillatory time dependence, i.e., a modification of the effective system chemical potential~\cite{Littlewood2006, Littlewood2007}. If $\Gamma$ is made sufficiently large, superconducting order will eventually be destroyed~\cite{Mitra2008}. Driving the valley $K_+$ with circularly polarised light $\sigma_+$, we find the following steady-state populations for momenta $\v{k} \in K_+$ and $\Delta_\pm = 0$
\begin{align}
    n_{-\v{k}}^{22} &= n_\text{F}^2 \, , \quad n_{\v{k}}^{22} = \frac{ n_\text{F}^2 +  \tilde{\Omega}_\v{k}^2(n_\text{F}^1/\gamma_2 + n_\text{F}^2/\gamma_1 )}{1  + \tilde{\Omega}_\v{k}^2(1/\gamma_2 + 1/\gamma_1)} \, , \\
    n_{-\v{k}}^{11} &= n_\text{F}^1 \, , \quad n_{\v{k}}^{11} = \frac{ n_\text{F}^1 + \tilde{\Omega}_\v{k}^2(n_\text{F}^1/\gamma_2 + n_\text{F}^2/\gamma_1)}{1 + \tilde{\Omega}_\v{k}^2(1/\gamma_2 + 1/\gamma_1)} \, ,
\end{align}
where $n_\text{F}^\alpha \equiv n_\text{F}(\xi_{\v{k}\alpha})$, $\tilde{\Omega}_\v{k}^2 \equiv \Omega^2 / (\epsilon_\v{k}^2 + \Gamma^2)$, and $\gamma_\alpha = \Gamma_\alpha / \Gamma$. That is, the valley $K_-$ is unaffected by the laser drive, while the populations in the valley $K_+$ are nonthermal.

The nonequilibrium gap equation \eqref{eqn:self-consistency-plus} may be written in matrix form as
\begin{equation}
    \begin{pmatrix}
        \Delta_+ \\
        \Delta_-
    \end{pmatrix}
    =
    \begin{pmatrix}
        V F_+ & v e^{i\phi} F_- \\
        v e^{-i\phi} F_+ & V F_-
    \end{pmatrix}
    \begin{pmatrix}
        \Delta_+ \\
        \Delta_-
    \end{pmatrix}
    \, .
\end{equation}
Including further scattering amplitudes simply increases the dimensionality of this matrix. To zeroth order in $\abs{V^\prime / V}$, the onset of superconductivity is determined solely by the behaviour of the two functions $F_\pm$ with increasing driving strength
\begin{equation}
    F_\pm \equiv -\frac{1}{N} \sum_{\v{k}\in K_\pm}   \frac{E_\v{k}}{E_\v{k}^2 + \Gamma^2} (1-n^{22}_{\v{k}} - n^{11}_{-\v{k}})
    \, .
\end{equation}
The induced population differences $1-n^{22}_{\v{k}} - n^{11}_{-\v{k}}$ for $\v{k} \in K_+$ and $\v{k} \in K_-$ have opposite sign, which is inherited by the functions $F_+$ and $F_-$. It is now also clear why interband pairing is more favourable with respect to intraband pairing: in equilibrium (at temperatures $T \ll \delta$), the population difference $1-n^{22}_{\v{k}}-n^{11}_{-\v{k}}$ vanishes, which means that the system is ``closer'' to a superconducting instability (i.e., the nonequilibrium populations $n^{\alpha\alpha}_\v{k}$ need only be modified slightly).
Substituting in our expressions for the steady-state values of the populations and defining $\bar{\gamma}^{-1} = \gamma_1^{-1} + \gamma_2^{-1}$, we arrive at
\begin{align}
    F_+ &= \frac{1}{2\gamma_2} \frac{-\mu}{\mu^2 + (\Gamma/2)^2} \int \mathrm{d}E \rho(E) \frac{\Omega^2}{\epsilon(E)^2 + \Omega^2/\bar{\gamma}+\Gamma^2}
    \, , \label{eqn:F-plus-expression} \\
    \quad F_- &= - \frac{\gamma_2}{\gamma_1} F_+
    \, ,
\end{align}
for temperatures $T \ll \delta$. The domain of integration extends over positive energies only. The density of states per unit cell $\rho(E)$ for hexagonal materials in the presence of a nonzero staggered chemical potential $\delta$, as in Eq.~\eqref{eqn:tight-binding-model}, can be evaluated exactly in terms of the corresponding gapless density of states $\rho_0$: $\rho(E)=(E/\tilde{E})\rho_0(\tilde{E})/4$, where $\tilde{E}=\sqrt{E^2-(\delta/2)^2}$~\cite{Pedersen2009} (the factor of 4 removes spin and valley degeneracy). Hereafter we will simplify to the symmetric choice $\gamma_1=\gamma_2$, in which case we find that $F_- = -F_+$. In the presence of a finite intervalley coupling $v=|V^\prime|$, the equation determining the onset of superconductivity reads 
\begin{equation}
    1 = (V^2 - v^2) F_+^2
    \, .
    \label{eqn:simplified-gap-equation}
\end{equation}
This expression represents our central result: \eqref{eqn:simplified-gap-equation} is insensitive to the sign of $V$, and therefore always has a solution as long as the driving is sufficiently strong.
This result is illustrated by the phase diagram in Fig.~\ref{fig:critical-coupling-a}. The two branches with opposite signs indicate that a solution is possible for both attractive and repulsive $V$.
As in thermal equilibrium, the normal state becomes unstable whenever such a superconducting solution exists.
For nonzero $V^\prime$, the critical $|V|$ does not tend to zero in the limit of large driving strengths, but instead saturates at a value $V_c = \pm |V^\prime|$. Evidently, then, it is desirable to have $|V^\prime|$ be as small as possible, which, as we have discussed, is automatically the case in real materials. In the extreme case $|V^\prime/V| = 1$, superconducting correlations can never develop, however strong the driving.

%%%%%%%%%%%%%%%%%%%%%%%%%%%%%%%%%%%%%%%%%%%%%%%%%%%%%%%%%%%%%%%%%%%%%%
%%%%%%%%%%%%%%%%%%%%%%%%%%%%%% KELDYSH %%%%%%%%%%%%%%%%%%%%%%%%%%%%%%%
%%%%%%%%%%%%%%%%%%%%%%%%%%%%%%%%%%%%%%%%%%%%%%%%%%%%%%%%%%%%%%%%%%%%%%

\textit{Keldysh Description}---Our argument in this letter was based on a mean field description and the Born--Markov approximation to describe the nonequilibrium steady state.
We show in the Supplementary Material that the latter assumption can be relaxed, and qualitatively similar results are obtained using a more complete Keldysh description of the problem.
At the expense of increasing the complexity of the theory, the benefits of the field-theoretic Keldysh description include: (i) quasiparticle states are populated thermally versus electron states, unlike in~\eqref{eqn:born-markov-time-evolution}, (ii) arbitrarily large damping $\Gamma$ may be included, and (iii) fluctuations about the mean field result may be included.
Most importantly, the two branches for $V_c$ with opposite sign seen in Fig.~\ref{fig:critical-coupling} are present in both approaches.

%%%%%%%%%%%%%%%%%%%%%%%%%%%%%%%%%%%%%%%%%%%%%%%%%%%%%%%%%%%%%%%%%%%%%%
%%%%%%%%%%%%%%%%%%%%%%%%%% IDEAL PARAMETERS %%%%%%%%%%%%%%%%%%%%%%%%%%
%%%%%%%%%%%%%%%%%%%%%%%%%%%%%%%%%%%%%%%%%%%%%%%%%%%%%%%%%%%%%%%%%%%%%%

\textit{Ideal parameters.}---The benefit of the simplified Born--Markov approach is that we are able to evaluate expressions explicitly, which allows us to make concrete statements about optimising the system parameters in order to minimise $V_c$. It is evident from \eqref{eqn:F-plus-expression} that the chemical potential should be chosen to be as close to $\pm\Gamma/2$ as possible. Assuming this optimal setup $\mu=-\Gamma/2$, $F_+$ in \eqref{eqn:F-plus-expression} evaluates approximately to
\begin{equation}
    F_+ \simeq \frac{A_c}{36 t}  \frac{\delta}{t}  \frac{(\Omega/\Gamma)^2}{\sqrt{1 + 4(\Omega/\Gamma)^2}}
    \, ,
\end{equation}
for $t \gtrsim \delta \gg \Omega$, $\Gamma$, having neglecting subleading corrections. $A_c = 3\sqrt{3}/2$ is the area of one unit cell. This expression suggests that one should (i) maximise the ratio $\delta/t$, which has the additional benefit of increasing the validity of our assumption about the driving pattern (see Fig.~\ref{fig:driving-pattern}), and (ii) minimise $\Gamma$ so that the physics of interest occurs at a lower laser power. It should be noted however that the magnitude of the gap also depends on $\Gamma$ (through $\Delta/\Gamma \sim \sqrt{\Omega/\Gamma}$ for $\Omega \gg \Gamma$) so a smaller damping rate also corresponds to a smaller superconducting gap.

%%%%%%%%%%%%%%%%%%%%%%%%%%%%%%%%%%%%%%%%%%%%%%%%%%%%%%%%%%%%%%%%%%%%%%
%%%%%%%%%%%%%%%%%%%%%%%%%%%%%% INTRABAND %%%%%%%%%%%%%%%%%%%%%%%%%%%%%
%%%%%%%%%%%%%%%%%%%%%%%%%%%%%%%%%%%%%%%%%%%%%%%%%%%%%%%%%%%%%%%%%%%%%%

\textit{Fast intraband relaxation.}---When the interband relaxation rate is slow with respect to the intraband rate $\Gamma_{12}$, the upper and lower bands (in the valley $K_+$) will separately equilibrate to quasithermal distributions with effective chemical potentials $\mu_2$ and $\mu_1$, respectively. These are determined by the driving strength in addition to the intraband relaxation rate and particle number conservation. (The gap equations for this regime are presented in the Supplementary Material.) 
The phase diagram for this limiting case is shown in
Fig.~\ref{fig:critical-coupling-b}, and is to be contrasted with its
counterpart, Fig.~\ref{fig:critical-coupling-a}.  Importantly, the two
branches for $V_c$ with opposite sign persist in this limit.
Quantitatively, however, the critical coupling strengths are
significantly smaller by virtue of a larger induced population
difference. Therefore, this regime where interband relaxation is slower than 
intraband relaxation, which is closer to the situation in real experiments, coincides 
with the case where superconductivity with repulsive interactions is most 
favorable. 

%%%%%%%%%%%%%%%%%%%%%%%%%%%%%%%%%%%%%%%%%%%%%%%%%%%%%%%%%%%%%%%%%%%%%%
%%%%%%%%%%%%%%%%%%%%%%%%%%%%%% OUTLOOK %%%%%%%%%%%%%%%%%%%%%%%%%%%%%%%
%%%%%%%%%%%%%%%%%%%%%%%%%%%%%%%%%%%%%%%%%%%%%%%%%%%%%%%%%%%%%%%%%%%%%%

\textit{Outlook.}---We have shown that interband superconducting correlations due to BCS pairing may develop in the presence of repulsive electronic interactions in two-dimensional materials which exhibit circular dichroism. Laser driving with circular polarisation $\sigma_+$ induces a nonthermal particle distribution in the $K_+$ valley, while the other valley remains unaffected. Hence, the nonequilibrium population deviation $1-n^{22}_{\v{k}}-n^{11}_{-\v{k}}$ that appears in the self-consistency condition has the opposite sign for the two valley--band subsystems. We demonstrated this mechanism for two limiting cases of dissipation.
Our results are of direct relevance to the monolayer transition metal
dichalcogenides, which satisfy the necessary criteria outlined in this
letter to potentially realise interband
superconductivity. 
They open the possibility of turning a range of insulating materials into 
superconductors at the flip of a switch.

%%%%%%%%%%%%%%%%%%%%%%%%%%%%%%%%%%%%%%%%%%%%%%%%%%%%%%%%%%%%%%%%%%%%%%
%%%%%%%%%%%%%%%%%%%%%%%%% ACKNOWLEDGEMENTS %%%%%%%%%%%%%%%%%%%%%%%%%%%
%%%%%%%%%%%%%%%%%%%%%%%%%%%%%%%%%%%%%%%%%%%%%%%%%%%%%%%%%%%%%%%%%%%%%%

\textit{Acknowledgements.}---The authors would like to thank Camille Aron for useful discussions. This work was supported in part by Engineering and Physical Sciences Research Council (EPSRC) Grants No.~EP/P034616/1 and No.~EP/M007065/1 (C.Ca.~and O.H.), by NSF grant No.~DMR-1733071 (G.G.), and by DOE Grant No.~DE- FG02-06ER46316 (C.Ch.). 
C.Ch.~acknowledges the kind hospitality of Trinity College, where this work started while visiting as a Fellow Commoner (of no common rate). 

%%%%%%%%%%%%%%%%%%%%%%%%%%%%%%%%%%%%%%%%%%%%%%%%%%%%%%%%%%%%%%%%%%%%%%
%%%%%%%%%%%%%%%%%%%%%%%%%%%% BIBLIOGRAPHY %%%%%%%%%%%%%%%%%%%%%%%%%%%%
%%%%%%%%%%%%%%%%%%%%%%%%%%%%%%%%%%%%%%%%%%%%%%%%%%%%%%%%%%%%%%%%%%%%%%

\bibliographystyle{aipnum4-1}
\bibliography{references}

\begin{thebibliography}{31}
\bibitem [{pauliblockNote()}]{pauliblockNote}%
  \BibitemOpen
  \bibinfo {note} {The populations of the valleys are altered in the presence
  of a nonzero superconducting gap, thus preventing Pauli
  blocking.}\BibitemShut {Stop}%
\end{thebibliography}

\end{document}

\title{Supplementary Material}

\maketitle

%%%%%%%%%%%%%%%%%%%%%%%%%%%%%%%%%%%%%%%%%%%%%%%%%%%%%%%%%%%%%%%%%%%%%%
%%%%%%%%%%%%%%%%%%%%%%%%%%%%% ASYMMETRY %%%%%%%%%%%%%%%%%%%%%%%%%%%%%%
%%%%%%%%%%%%%%%%%%%%%%%%%%%%%%%%%%%%%%%%%%%%%%%%%%%%%%%%%%%%%%%%%%%%%%

\section{Asymmetry}

In this section we describe how Fig.~2 in the main text was calculated. The definition of the so-called degree of circular polarisation, $\eta(\v{k})$, which quantifies the asymmetry between absorption of the two different circular polarisations $\sigma_\pm$ is~\cite{Yao2008, Cao2012}
\begin{equation}
    \eta(\v{k}) = \frac{\abs*{\mc{P}_+^{21}(\v{k})}^2 - \abs*{\mc{P}_-^{21}(\v{k})}^2}{\abs*{\mc{P}_+^{21}(\v{k})}^2 + \abs*{\mc{P}_-^{21}(\v{k})}^2}
    \, .
\end{equation}
The matrix elements $\mc{P}^{21}_\pm(\v{k}) = \mel*{\psi_{2\v{k}}}{p_\pm}{\psi_{1\v{k}}}$ with $p_\pm = p_x \pm i p_y$ describe vertical transitions between the conduction ($\alpha = 2$) and valence ($\alpha = 1$) bands in momentum space, induced by driving the system with light of circular polarisation $\sigma_\pm$. Let us analyse the problem of evaluating $\eta(\v{k})$ for the tight-binding Hamiltonian
\begin{equation}
    H(\v{k}) = 
    \begin{pmatrix}
        \delta/2 & h(\v{k}) \\
        h^*(\v{k}) & -\delta/2
    \end{pmatrix}
    \, ,
\end{equation}
which describes hexagonal materials in the presence of a staggered chemical potential $\pm \delta/2$. Let us denote the corresponding (normalised) eigenvectors by $\Psi_\alpha(\v{k}) = u_\alpha(\v{k})\psi_\v{k}^a + v_\alpha(\v{k})\psi_\v{k}^b$ where $\alpha = 1, 2$ is the band index and $a, b$ correspond to the two sublattices. The basis states $\psi_\v{k}^x$ are given in position space by
\begin{equation}
    \bra{\v{r}}\ket{\psi_\v{k}^{x}} = \frac{1}{\sqrt{N}} \sum_{i} e^{-\v{k}\cdot \v{R}_i^x} w(\v{r} - \v{R}_i^x)
    \, ,
\end{equation}
where $\{\v{R}_i^x\}$ denote atomic positions on sublattice $x=a,b$, the number of unit cells is $N$, and $w(\v{r})$ is the appropriate Wannier state. In terms of these basis states, the matrix elements are given by
\begin{align}
    \mc{P}_\pm^{21}(\v{k}) &= u^*_2 v_1 \mel*{\psi_\v{k}^a}{p_\pm}{\psi_\v{k}^b} + v^*_2 u_1 \mel*{\psi_\v{k}^b}{p_\pm}{\psi_\v{k}^a}
     \\
    &= u^*_2 v_1 \mel*{\psi_\v{k}^a}{p_\pm}{\psi_\v{k}^b} + v^*_2 u_1 \mel*{\psi_\v{k}^a}{p_\mp}{\psi_\v{k}^b}^* \, .
\end{align}
The diagonal contributions (i.e., $\mel*{\psi_\v{k}^x}{p_\pm}{\psi_\v{k}^x}$) to the above expression vanish identically by parity and so have already been dropped. Hence, we need only calculate two matrix elements, $\mel*{\psi_\v{k}^a}{p_\pm}{\psi_\v{k}^b}$, in order to find $\eta(\v{k})$. One may show that, on a hexagonal lattice,
\begin{equation}
    \mel*{\psi_\v{k}^a}{p_\pm}{\psi_\v{k}^b} \propto (1 \pm i\sqrt{3})e^{i\v{k}\cdot \v{d}_1} + (1 \mp i\sqrt{3})e^{i\v{k}\cdot \v{d}_2} - 2e^{i\v{k}\cdot \v{d}_3} + \ldots
    \, ,
\end{equation}
where $\v{d}_{1,2}=\tfrac{a}{2}\v{y} \pm \tfrac{\sqrt{3}a}{2}\v{x}$ and $\v{d}_3 = -a\v{y}$ are the vectors connecting nearest neighbours~\cite{Castro2009}.
The dots represent terms beyond nearest-neighbour contributions. Here we neglect them, consistent with the spirit of the original tight-binding model.
The constant of proportionality is irrelevant for a calculation of $\eta(\v{k})$, and depends on the details of $w(\v{r})$. Collecting the above results, one may deduce an explicit expression for the asymmetry. For any nonzero bandgap $\delta$, the selection rule is exact at the points $K_\pm$ which correspond to the centres of the valleys. Near $K_\pm$, we have that $\eta(\v{K}_\pm + \boldsymbol{\kappa}) \mp 1 \propto (t/\delta) \kappa^2$, suggesting that maximising $\delta / t$ makes the asymmetry more prominent. 

%%%%%%%%%%%%%%%%%%%%%%%%%%%%%%%%%%%%%%%%%%%%%%%%%%%%%%%%%%%%%%%%%%%%%%
%%%%%%%%%%%%%%%%%%%%%%%%%%%% BORN MARKOV %%%%%%%%%%%%%%%%%%%%%%%%%%%%%
%%%%%%%%%%%%%%%%%%%%%%%%%%%%%%%%%%%%%%%%%%%%%%%%%%%%%%%%%%%%%%%%%%%%%%

\section{Born--Markov Derivation}

In this section we elucidate and add detail to our results relating to the Born--Markov analysis of the nonequilibrium steady state. In particular, we write down explicitly the appropriate equations of motion for system populations and correlators that determine the nonequilibrium populations in the long-time limit presented in the main text.

The time evolution of our system is governed approximately by the equation of motion~\cite{Puri2001}
\begin{equation}
    \dv{\rho_\text{S}}{t} = -i [H_\text{S}, \rho_\text{S}] + \sum_{\lambda} \Gamma_\lambda \Big\{ n_\text{F}(\xi_\lambda) \mc{D}[c_\lambda^\dagger]\rho_\text{S} + [1-n_\text{F}(\xi_\lambda)] \mc{D}[c_\lambda] \rho_\text{S} \Big\} \, ,
    \label{supp:eqn:born-markov-eom}
\end{equation}
for the system density matrix $\rho_\text{S}(t) = \Tr_\text{B}\rho(t)$, having traced out the bath degrees of freedom.
The equations of motion for correlators and populations are given by $\partial_t\ev*{O(t)}  = \Tr O \partial_t \rho(t)$, with $\partial_t \rho$ given by the above expression. In particular, in the frame corotating with the laser at frequency $\omega_0$, the relevant populations and correlators that form a closed set under time evolution are
\begin{align}
    \dv{t} n_\v{k}^{11} &= -i \Omega_\v{k} (n_\v{k}^{12} - n_\v{k}^{21}) - i \Delta_{-\v{k}} s_{-\v{k}}^{21} + i\bar{\Delta}_{-\v{k}} \bar{s}_{-\v{k}}^{21} - 2\Gamma_1 [n_\v{k}^{11} - n_\text{F}(\xi_{\v{k}1})] \, , \label{supp:eqn:born-markov-n11} \\[0.15cm]
    \dv{t} n_\v{k}^{22} &= +i \Omega_\v{k} (n_\v{k}^{12} - n_\v{k}^{21}) - i \Delta_\v{k} s_\v{k}^{21} + i\bar{\Delta}_\v{k} \bar{s}_\v{k}^{21} - 2\Gamma_2 [n_\v{k}^{22} - n_\text{F}(\xi_{\v{k}2})] \, , \label{supp:eqn:born-markov-n22} \\[0.15cm]
    \dv{t} n_\v{k}^{21} &= i\epsilon_\v{k} n_\v{k}^{21} - i \Omega_\v{k} (n_\v{k}^{22} - n_\v{k}^{11}) - \Gamma n_\v{k}^{21}  \, , \label{supp:eqn:born-markov-n21} \\[0.15cm]
    \dv{t} s_{\v{k}}^{21} &= i E_\v{k} s_\v{k}^{21} + i\bar{\Delta}_\v{k} (1 - n_\v{k}^{22}  - n_{-\v{k}}^{11} ) - \Gamma s_\v{k}^{21} \label{supp:eqn:born-markov-s21}
    \, ,
\end{align}
where $E_\v{k} = \xi_{\v{k}1} + \xi_{\v{k}2}$, $\epsilon_\v{k} = \xi_{\v{k}2}-\xi_{\v{k}1} - \omega_0$, $\xi_{\v{k}\alpha} = E_{\v{k}\alpha} - \mu$ and $\Gamma = \Gamma_1 + \Gamma_2$. Recall the definitions $n_\v{k}^{\alpha\beta}(t) = \ev*{c^\dagger_{\v{k}\alpha} c^{\phantom{\dagger}}_{\v{k}\beta}}_{\rho(t)}$ and $s_\v{k}^{\alpha\beta}(t) = \ev*{c^\dagger_{\v{k}\alpha} c^{\dagger}_{-\v{k}\beta}}_{\rho(t)}$ used in the main text. These equations are a generalisation of those given in Ref.~\cite{Goldstein2015}. In order to derive these equations, we have made use of the identity $\Tr O\mc{D}[X]\rho(t) = \Tr\comm*{X^\dagger}{O}X\rho(t) + \Tr X^\dagger \comm*{O}{X}\rho(t) = \ev*{\comm*{X^\dagger}{O}X}_{\rho(t)} + \ev*{X^\dagger \comm*{O}{X}}_{\rho(t)}$ repeatedly, where the Lindbladian dissipator $\mc{D}$ is defined as $\mc{D}[X]\rho = (2 X\rho X^\dagger - X^\dagger X\rho - \rho X^\dagger X )/2$. We then look for steady-state solutions to the set of equations \eqref{supp:eqn:born-markov-n11}--\eqref{supp:eqn:born-markov-s21}. The last equation, \eqref{supp:eqn:born-markov-s21}, can be rearranged to find
\begin{equation}
    s_\v{k}^{21} = -\frac{\bar{\Delta}_\v{k}(1 - n_\v{k}^{22}  - n_{-\v{k}}^{11})}{E_\v{k} + i\Gamma}
    \, .
\end{equation}
Self-consistency is then obtained by solving $\bar{\Delta}_\v{k} = (1/N) \sum_{\v{k}^\prime} V_{\v{k}\v{k}^\prime} \bar{\Delta}_{\v{k}^\prime} \Re ( s^{21}_{\v{k}^\prime} / \bar{\Delta}_{\v{k}^\prime} )$. The real part in this expression requires some further explanation. Note that the equations \eqref{supp:eqn:born-markov-n11}--\eqref{supp:eqn:born-markov-s21} permit a steady-state solution for $n_\v{k}^{\alpha \beta}$, and an oscillating solution for $\Delta_\v{k}(t)$ of the form $\Delta_\v{k}(t) = \Delta_\v{k}^0 e^{i\nu_\v{k} t}$~\cite{Littlewood2006, Littlewood2007}. Physically, this corresponds to a shift in the effective system chemical potential due to interactions between the system and the bath. A full solution would correspond to solving both the real and imaginary parts of the gap equation, which fix both the magnitude of the gap and the effective chemical potential of the system.
In our simplified approach, we take the damping $\Gamma$ to be small such that the effective system chemical potential approximately equals the bath chemical potential (i.e., $\nu_\v{k} \simeq 0$ in the ansatz). After making this simplification, we are left with just one equation and one free parameter (for each $\Delta_\v{k}$): the magnitude of the gap. In fact, a complete solution must also include a modification of the disspipative term in~\eqref{supp:eqn:born-markov-eom} in order to thermally populate the quasiparticle (as opposed to electronic) states. All of these subtleties are fixed in the Keldysh calculation which we present in the next section.

Solving the self-consistency condition requires eliminating $n_\v{k}^{21}$ from \eqref{supp:eqn:born-markov-n11}--\eqref{supp:eqn:born-markov-n21} in order to find the nonequilibrium populations of the upper and lower bands, $n_\v{k}^{11}$ and $n_\v{k}^{22}$, subject (in general) to nonzero drive $\Omega_\v{k}$ and nonzero superconducting correlations $\Delta_\v{k}$. In the simplified case $V^\prime = 0$, the two valley--band subsystems decouple and we may consider the equations for $\Delta_+$ and $\Delta_-$ separately. Assuming the simplified driving pattern outlined in the main text, and that $\Delta_-=0$, $\Delta_+=\Delta$, we arrive at
\begin{equation}
    1 - n_\v{k}^{22}  - n_{-\v{k}}^{11} = -\frac{\gamma_1^2 \tilde{\Omega}_\v{k}^2 }{ \gamma_1[\gamma_1\gamma_2 + (\gamma_1 + \gamma_2)\tilde{\Delta}_\v{k}^2] +[\gamma_1(\gamma_1 + \gamma_2) + (2\gamma_1 + \gamma_2)\tilde{\Delta}_\v{k}^2]\tilde{\Omega}_\v{k}^2 } \, ,
    \label{supp:eqn:population-difference-complete}
\end{equation}
for $\v{k} \in K_+$, where $\tilde{\Omega}_\v{k}^2=\Omega^2/(\epsilon_\v{k}^2 + \Gamma^2)$ and $\tilde{\Delta}_\v{k}^2 = \abs{\Delta}^2/(E_\v{k}^2 + \Gamma^2)$, for temperatures $T \ll \delta$ such that $n_\text{F}(\xi_{\v{k}1}) \simeq 1$ and $n_\text{F}(\xi_{\v{k}2}) \simeq 0$. If instead $\Delta_+ = 0$, $\Delta_-=\Delta$, then
\begin{equation}
    1 - n_\v{k}^{22}  - n_{-\v{k}}^{11} = +\frac{\gamma_2^2 \tilde{\Omega}_\v{k}^2  }{\gamma_2[\gamma_1\gamma_2 + (\gamma_1 + \gamma_2)\tilde{\Delta}_\v{k}^2 ] + [\gamma_2(\gamma_1 + \gamma_2)+(2\gamma_2+\gamma_1)\tilde{\Delta}_\v{k}^2]\tilde{\Omega}_\v{k}^2 }
    \, ,
    \label{supp:eqn:population-difference-complete-minus}
\end{equation}
for $\v{k} \in K_-$, i.e., the same form as~\eqref{supp:eqn:population-difference-complete} but with $\gamma_1 \leftrightarrow \gamma_2$ and, crucially, a sign flip. One must drive the system in order to induce a population difference of the form $1 - n_\v{k}^{22}  - n_{-\v{k}}^{11} \neq 0$.

In the most general case, when both $\Delta_+$ and $\Delta_-$ are considered to be nonzero, which becomes necessary if we consider nonzero intervalley coupling $V^\prime \neq 0$, the nonequilibrium populations of the states are found by solving the matrix equation
\begin{equation}
    \begin{pmatrix}
        \gamma_2 + \abs*{\tilde{\Delta}_+}^2 + \tilde{\Omega}^2 & \abs*{\tilde{\Delta}_+}^2 & -\tilde{\Omega}^2 & 0 \\
        \abs*{\tilde{\Delta}_+}^2 & \gamma_1 + \abs*{\tilde{\Delta}_+}^2 & 0 & 0 \\
        -\tilde{\Omega}^2 & 0 & \gamma_1 + \abs*{\tilde{\Delta}_-}^2 + \tilde{\Omega}^2 & \abs*{\tilde{\Delta}_-}^2 \\
        0 & 0 & \abs*{\tilde{\Delta}_-}^2 & \gamma_2 + \abs*{\tilde{\Delta}_-}^2
    \end{pmatrix}
    \begin{pmatrix}
        n_\v{k}^{22} - 1/2 \\[0.15cm]
        n_{-\v{k}}^{11} - 1/2 \\[0.15cm]
        n_{\v{k}}^{11} - 1/2 \\[0.15cm]
        n_{-\v{k}}^{22} - 1/2
    \end{pmatrix}
    =
    \begin{pmatrix}
        \gamma_2(n_\text{F}^2 - 1/2) \\[0.15cm]
        \gamma_1(n_\text{F}^1 - 1/2) \\[0.15cm]
        \gamma_1(n_\text{F}^1 - 1/2) \\[0.15cm]
        \gamma_2(n_\text{F}^2 - 1/2)
    \end{pmatrix}
    \, .
\end{equation}

For $\Delta=0$, the expressions \eqref{supp:eqn:population-difference-complete} and \eqref{supp:eqn:population-difference-complete-minus} reduce to the ones given in the main text which describe the onset of superconductivity in our system. In particular, summing over momenta and converting the summation into an integral over energy using the density of states per unit cell $\rho(E)$, we arrive at the following expression for the functions $F_\pm$
\begin{align}
    F_+ &= \frac{1}{\gamma_2} \left( -\frac{\mu}{\mu^2 + \Gamma^2} \right) \int_{E>0} \mathrm{d}E \rho(E) \frac{\Omega^2}{\epsilon(E)^2 + \Omega^2/\bar{\gamma}+\Gamma^2}
    \, , \label{supp:eqn:F-plus-expression} \\
    \quad F_- &= - \frac{\gamma_2}{\gamma_1} F_+
    \, .
\end{align}
The integral in~\eqref{supp:eqn:F-plus-expression} may be evaluated in the limit of small $\Omega, \Gamma \ll \delta \lesssim t$ as follows. First, we make use of the approximate density of states
\begin{equation}
    \rho(E) \simeq \frac{A_c\abs*{E}}{2\pi  v_\text{F}^2} \Theta(\abs*{E} - \delta/2)
    \, ,
    \label{supp:eqn:approx-dos}
\end{equation}
where $v_\text{F} = 3 t/2$ and $\Theta(x)$ is the Heaviside step function. This expression may be derived using the approximate dispersion $E_\v{k} = t\sqrt{|h(\v{k})|^2+(\delta/2)^2}$ valid close to the centres of the valleys, or by expanding the exact density of states given in the main text. We then plug~\eqref{supp:eqn:approx-dos} into~\eqref{supp:eqn:F-plus-expression} and make use of the standard integral
\begin{align}
    \int_{x_0}^{x_1} \mathrm{d}x \, \frac{x}{(x-x_0)^2 + \eta^2} &= \frac{x_0}{\eta} \arctan\left( \frac{x_1 - x_0}{\eta}  \right) + \frac{1}{2} \log\left[ 1 + {\left( \frac{x_1-x_0}{\eta} \right)}^2 \right] \\
    &\simeq \frac{\pi x_0}{2\eta} \, ,
\end{align}
where in the second line we have taken the limit of small $\eta$ ($\eta \ll x_0$ and $\eta \ll x_1 - x_0$), i.e., $\Omega, \Gamma \ll \delta$. Evidently, to lowest order in $1/\eta$, the cutoff $x_1$ is irrelevant, with the dominant contribution coming from a region of width $\sim \eta$ around $x_0$. Using the correspondences $x_0 \to \delta/2$, $\eta \to \sqrt{\Omega^2 + \Gamma^2/4}$, we arrive at the result stated in the main text for $F_\pm$, from which one can deduce expressions for $V_c$. Higher order corrections may be included by choosing the cutoff $x_1$ such that the total number of states is preserved.

%%%%%%%%%%%%%%%%%%%%%%%%%%%%%%%%%%%%%%%%%%%%%%%%%%%%%%%%%%%%%%%%%%%%%%
%%%%%%%%%%%%%%%%%%%%%%%%%%%%%% KELDYSH %%%%%%%%%%%%%%%%%%%%%%%%%%%%%%%
%%%%%%%%%%%%%%%%%%%%%%%%%%%%%%%%%%%%%%%%%%%%%%%%%%%%%%%%%%%%%%%%%%%%%%

\section{Keldysh Description}

The majority of the approximations (both controlled and uncontrolled) made in the Born--Markov approach described in the previous section can be made more rigorous using a more careful analysis of the problem within the Keldysh formalism. We also relax the assumption of spinlessness and endow the electrons with a spin degree of freedom for this analysis. We assume a spin singlet interaction between electrons of the form
\begin{equation}
    H_\text{int} = \frac{1}{N} \sum_{\v{k}, \v{k}^\prime} V_{\v{k}\v{k}^\prime} \Big( c^\dagger_{\v{k}2\uparrow} c^\dagger_{-\v{k}1\downarrow} - c^\dagger_{\v{k}2\downarrow} c^\dagger_{-\v{k}1\uparrow} \Big) \Big( c_{-\v{k}^\prime 1\downarrow}c_{\v{k}^\prime 2 \uparrow} -  c_{-\v{k}^\prime 1\uparrow}c_{\v{k}^\prime 2 \downarrow} \Big)
    \, ,
    \label{supp:eqn:k-dependent-interaction}
\end{equation}
i.e., the scattering of interband spin singlets~\cite{Tahir-Kheli1998}.

The bath degrees of freedom can straightforwardly be integrated out to give the components (retarded `R', advanced `A' and Keldysh `K') of the self-energy $\Sigma_\lambda$
\begin{align}
    \Sigma_{\lambda}^\text{R}(\omega) &= -i\Gamma_{\lambda}(\omega) - \delta_{\lambda}(\omega) \, , \\
    \Sigma_{\lambda}^\text{A}(\omega) &= \Sigma_\lambda^\text{R}(\omega)^* \, , \\
    \Sigma_{\lambda}^\text{K}(\omega) &= -2i\Gamma_{\lambda}[1-2n_\text{F}(\omega, \mu)] \, ,
\end{align}
where the imaginary terms give rise to dissipation, and the real parts lead to a renormalisation (Lamb shift) of the bandstructure $E_\lambda \to E_\lambda + \delta_\lambda$. The multi-index $\lambda$ that labels the system modes now also includes spin: $\lambda = (\v{k}, \alpha, \sigma)$. Both the real and imaginary parts can be written in terms of the spectral densities of the baths $J_{\lambda}(\omega) = \sum_{n}\abs*{t_{\lambda n}}^2 \delta(\omega - \omega_{\lambda n})$:
\begin{align}
    \Gamma_\lambda(\omega) &= \pi J_\lambda(\omega) \, , \\ 
    \delta_\lambda(\omega) &= \mathcal{P} \!\! \int \mathrm{d}\omega^\prime \frac{J_\lambda(\omega^\prime)}{\omega^\prime - \omega} \, .
\end{align}
The symbol $\mc{P}$ denotes the principal value of the integral.
In the main text, we assumed $t_{\lambda n} = t_\lambda \, \forall n$ so that $J_\lambda(\omega) = \abs*{t_\lambda}^2 \nu_\lambda (\omega)$, where $\nu_\lambda(\omega)$ is the bath density of states corresponding to system mode $\lambda$.
For simplicity, we ignore the energy renormalisation, $\delta_{\lambda} = 0$, and neglect the frequency and momentum dependence of the dissipation parameters $\Gamma_{\lambda}(\omega) = \Gamma_\alpha$, corresponding to a bath with a flat spectral density, $J_\lambda = \text{const}$.

In order to proceed with the calculation, we transform into the frame corotating with the laser drive, and make the rotating wave approximation, i.e., neglecting the nonresonant terms that rotate at angular frequency $2\omega_0$.
In the main text, we assumed a Hamiltonian that includes only the resonant terms from the outset.
Without this approximation, the order parameter would not be a time-independent quantity, and we would need to include higher harmonics rotating at $e^{2n i \omega_0  t}$ ($n \in \mathbb{Z}$) in our ansatz. Importantly, in the rotating frame, we must implement a frequency shift in the Keldysh component $\Sigma_\lambda^\text{K}(\omega)$ to account for the fact that, in the lab frame in the absence of driving, the bath states are populated thermally, i.e.,
\begin{equation}
    \Sigma_{\lambda}^\text{K}(\omega) = -2{i}\Gamma_\alpha \tanh\left[\tfrac12 \beta (\omega + \tfrac12 (-1)^\alpha \omega_0 - \mu ) \right]
    \, ,
    \label{supp:eqn:keldysh-component-self-energy}
\end{equation}
where $T=1/\beta$ and $\mu$ are the temperature and chemical potential of the bath, respectively.

The quartic interaction term~\eqref{supp:eqn:k-dependent-interaction} can be decoupled in the standard way using a Hubbard Stratonovich transformation~\cite{Altland2010}. The resulting action for the gap function $\Delta_\v{k}$ and for the system fermions $\psi$ is given by
\begin{equation}
    S[\boldsymbol{\psi}_\v{k}, \bar{\boldsymbol{\psi}}_\v{k}, \Delta_\v{k}, \bar{\Delta}_\v{k}] = \int_{-\infty}^{\infty} \mathrm{d}t \, \sum_{\v{k}} \bar{\boldsymbol{\psi}}_\v{k} (\hat{\mathcal{G}}^{-1}_{0\v{k}} - \hat{\Sigma}_\v{k} + \hat{\Delta}_\v{k} + \hat{\Delta}_\v{k}^\dagger  ) \boldsymbol{\psi}_\v{k} + N\sum_{\v{k}\v{k}^\prime} \bar{\Delta}_\v{k} (V^{-1})_{\v{k}\v{k}^\prime} \Delta_{\v{k}^\prime}
    \, ,
    \label{supp:eqn:action}
\end{equation}
where $\hat{\mathcal{G}}_{0\lambda}$ is the system Green's function that includes nonzero driving. We have also defined $\hat{\Delta}_\v{k} = \sum_{a,\tau} \Delta_{\tau\v{k}}^a(t)\hat{\gamma}_{a,\tau}$, where $a$ is the Keldysh index $a=\text{q},\text{cl}$ and $\tau = \pm 1$, and we have introduced the 8-component spinor $\boldsymbol{\psi}_\v{k}$ in $\text{Keldysh}\otimes\text{Nambu}\otimes\text{band}$ space (hats denote $8 \!\times\! 8$ matrices in this space).

The most general gap equation corresponding to the saddle point of the action \eqref{supp:eqn:action} is given by varying the action with respect to the quantum component of the order parameter, which leads to
\begin{equation}
    \Delta_\v{k}(t) = -\frac{{1}}{2iN}  \sum_{\v{k}^\prime, \pm} V_{\v{k} \v{k}^\prime} \Tr \mc{G}(\pm \v{k}^\prime; t, t) \gamma_{\text{q}\pm}^\dagger
    \, ,
    \label{supp:eqn:keldysh-gap-eqn}
\end{equation}
where $\mc{G}(\v{k}; t, t)$ is the full time-dependent system Green's function, which includes the effects of driving, interactions with the bath through $\Sigma_\lambda(\omega)$, and nonzero $\{\Delta_\v{k}\}$. We have here redefined $ \Delta_\v{k}(t) \equiv \Delta_\v{k}^\text{cl}(t)$. The equation~\eqref{supp:eqn:keldysh-gap-eqn} is simply the nonequilibrium, multiband generalisation of the familiar BCS self-consistency condition $\Delta_\v{k} = (1/N)\sum_{\v{k}^\prime} V_{\v{k}\v{k}^\prime} \ev*{c_{-\v{k}^\prime\downarrow} c_{\v{k}^\prime\uparrow}}$. The equations obtained by varying the action with respect to the classical components of the gap function are automatically satisfied by setting $\Delta_\v{k}^\text{q}=0$, $\forall \v{k}$.

The gap equation \eqref{supp:eqn:keldysh-gap-eqn} can then be solved in terms of the time-dependent ansatz $\Delta_\v{k}(t) = e^{-2i(\mu+\delta\mu_\v{k})t} \Delta_\v{k}$. In the limit of zero damping $\Gamma_\alpha \to 0^+$ (keeping the ratio $\Gamma_1 / \Gamma_2$ fixed), the bath induces an effective chemical potential of $\mu$ in the system so that $\delta\mu_\v{k} = 0$. The gap equation \eqref{supp:eqn:keldysh-gap-eqn} is then purely real, and the free parameters are the magnitudes of the gap parameters $\Delta_\v{k}$. If the damping $\Gamma_\alpha$ remains finite, one must clearly solve twice as many equations: both the real and imaginary part of~\eqref{supp:eqn:keldysh-gap-eqn}, with twice as many parameters: the magnitudes of the order parameters $\{\abs*{\Delta_\v{k}}\!\}$ and the shifts in the effective chemical potential $\{ \delta\mu_\v{k} \}$.

Since the two valleys $K_\pm$ are separated by a large momentum, the amplitudes for intra- and inter-valley scattering are expected to be vastly different. We account for this by introducing the two scattering amplitudes $V$ and $V^\prime$, satisfying $\abs*{V^\prime} \ll \abs*{V}$, which represent some average scattering amplitude for intra- and inter-valley scattering processes, respectively. All other momentum-dependence of the scattering amplitudes is neglected.

\begin{figure}
    \centering
    \includegraphics[width=0.5\linewidth]{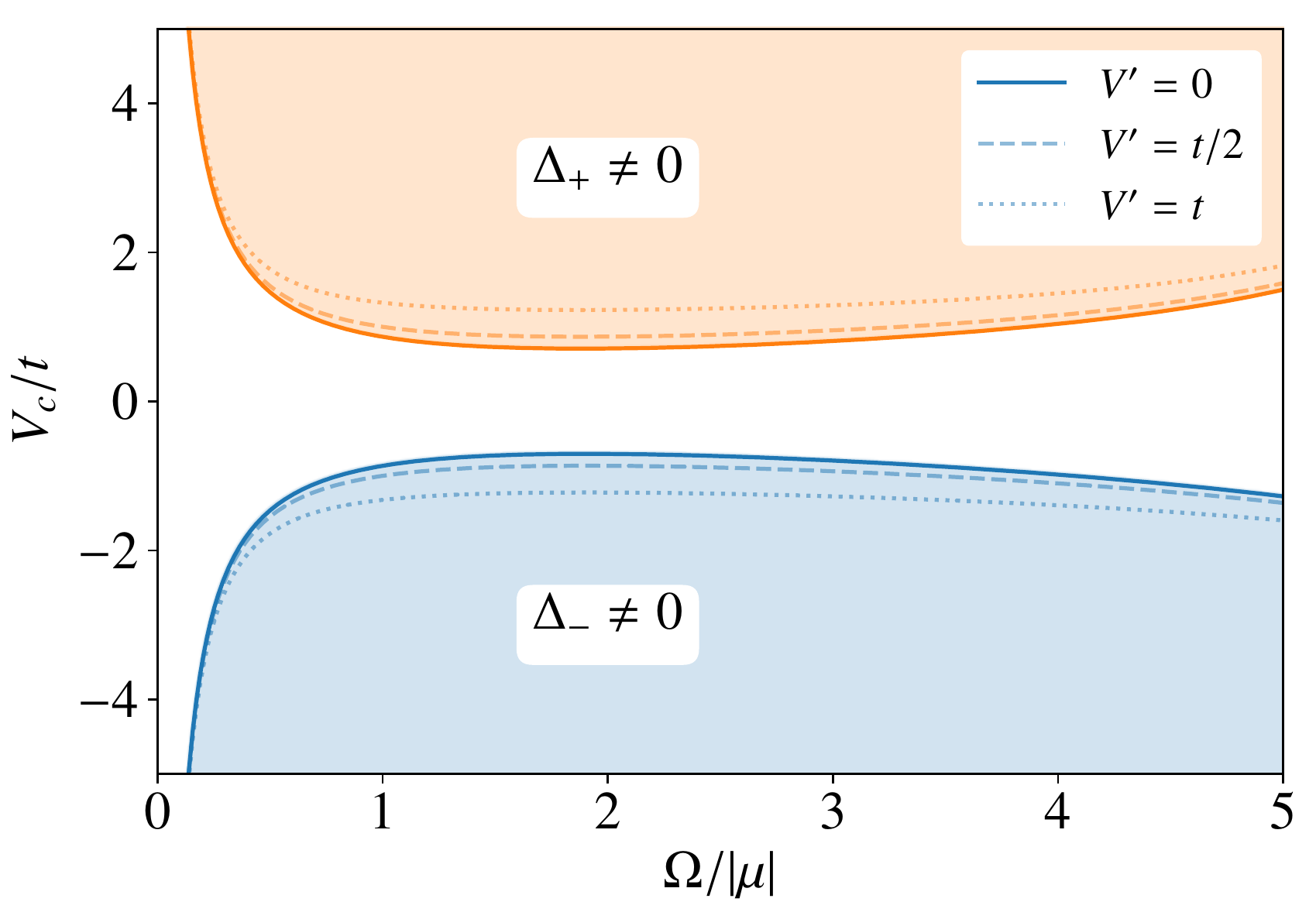}
    \caption{Critical coupling $V_c$ as predicted by the Keldysh description, with the shaded regions corresponding to a nonzero superconducting order parameter. As in the main text, the critical curve $V_c$ has two branches with opposite signs implying that superconducting correlations can devlep irrespective of the sign of the electronic interactions. The perfect symmetry $F_-=-F_+$ breaks down for $\Omega \gtrsim |\mu|$, and the curves are no longer monotonic functions of the driving strength. The parameters used for the plot are $\delta/t = 5$ and $\mu/t=-2.5\times 10^{-4}.$}
    \label{fig:keldysh-Vc}
\end{figure}

Having restricted our attention to only these two relevant scattering amplitudes, we need only retain two gap parameters $\Delta_\v{k} = \Delta_\pm$ for $\v{k} \in K_\pm$. These two parameters characterise the extent of pairing across the two valleys, as indicated schematically in Fig.~1 in the main text. The gap equation~\eqref{supp:eqn:keldysh-gap-eqn} can then be written in matrix form
\begin{equation}
    \frac{1}{N} \sum_{\v{k} \in K_+} \int \frac{\mathrm{d}\omega}{2\pi} 
    \begin{pmatrix}
        V  f^+_{\v{k}}(\omega) &  V^\prime  f^-_{\v{k}}(\omega) \\
        \bar{V}^\prime f^+_{\v{k}}(\omega) &   V f^-_{\v{k}}(\omega)
    \end{pmatrix}
    \begin{pmatrix}
        \Delta_+ \\
        \Delta_-
    \end{pmatrix}
    =
    \begin{pmatrix}
        \Delta_+ \\
        \Delta_-
    \end{pmatrix}
    \, ,
    \label{supp:eqn:simplified-matrix-gap-eqn}
\end{equation}
where the sums are evaluated over momenta in one of the valleys only (say $\v{k} \in K_+$), and
\begin{equation}
    f^\tau_\v{k}(\omega) = \frac{i}{2\Delta_\tau} \sum_{\tau^\prime=\pm 1}\Tr \mc{G}_{\tau^\prime}(\v{k}, \omega) \gamma^\dagger_{\text{q},\tau\tau^\prime}
    \, .
    \label{supp:eqn:f_k-definition}
\end{equation}
Although it is not immediately apparent from their definition~\eqref{supp:eqn:f_k-definition}, the functions $f_\v{k}^\tau(\omega)$ depend on the gap parameters separately through $\abs*{\Delta_+}^2$ and $\abs*{\Delta_-}^2$ [i.e., there are no mixed terms such as $\Re(\Delta_+\Delta_-^*)$]. The Fourier-transformed inverse Green's functions are given by
\begin{equation}
    \mc{G}^{-1}_\tau(\v{k}, \omega) = 
    \begin{pmatrix}
        [\mc{G}^\text{R}_\tau]^{-1} & -\Sigma^\text{K} \\
        0 & [\mc{G}^\text{A}_\tau]^{-1}
    \end{pmatrix}
    \, ,
    \label{supp:eqn:Keldysh-greens-function-momentum-space}
\end{equation}
where the retarded components of the full Green's function~\eqref{supp:eqn:Keldysh-greens-function-momentum-space} may be expressed in $\text{Nambu} \otimes \text{band}$ space as
\begin{widetext}
\begin{align}
    [\mathcal{G}^\text{R}_+]^{-1}(\v{k}, \omega) = 
    \begin{pmatrix}
        \omega - \xi_{1\v{k}} + i\Gamma_1 & \Omega & 0 & \Delta_- \\
        \Omega & \omega - \xi_{2\v{k}} +i\Gamma_2 & \Delta_+ & 0 \\
        0 & \bar{\Delta}_+ & \omega + \xi_{1,-\v{k}} +i\Gamma_1 & 0 \\
        \bar{\Delta}_- & 0 & 0 & \omega + \xi_{2,-\v{k}} + i \Gamma_2
    \end{pmatrix} \, , \label{supp:eqn:GR-plus} \\
    [\mathcal{G}^\text{R}_-]^{-1}(\v{k}, \omega) = 
    \begin{pmatrix}
        \omega - \xi_{1\v{k}} + i\Gamma_1 & 0 & 0 & \Delta_+ \\
        0 & \omega - \xi_{2\v{k}} +i\Gamma_2 & \Delta_- & 0 \\
        0 & \bar{\Delta}_- & \omega + \xi_{1,-\v{k}} +i\Gamma_1 & - \Omega \\
        \bar{\Delta}_+ & 0 & -\Omega & \omega + \xi_{2,-\v{k}} + i \Gamma_2
    \end{pmatrix}
    \, , \label{supp:eqn:GR-minus}
\end{align}
\end{widetext}
and $\xi_{\alpha\v{k}}=\epsilon_{\alpha\v{k}} - \mu$. The Keldysh component of the inverse Green's function is given by the self energy~\eqref{supp:eqn:keldysh-component-self-energy}. The Keldysh component of the Green's function is then given by $\mc{G}_\tau^\text{K}(\v{k}, \omega) = \mc{G}^\text{R}_\tau\Sigma^\text{K}\mc{G}^\text{A}_\tau$. When computing the summations over momentum, we convert to an integral over energy with the exact density of states for gapped hexagonal materials stated in the main text. Alternatively, if $\delta \ll t$, one may use the approximate density of states~\eqref{supp:eqn:approx-dos} with a cutoff energy chosen to preserve the total number of states.

In the limiting case $V^\prime=0$, the gap equation~\eqref{supp:eqn:simplified-matrix-gap-eqn} does not mix the components $\Delta_+$ and $\Delta_-$, so that the matrix equation reduces to two separate equations:
\begin{equation}
    \left[1 - V F_\pm \right]\Delta_\pm = 0 \, , \qquad\qquad F_\pm \equiv \frac{1}{N} \sum_{\v{k}\in K_+} \int \frac{\mathrm{d}\omega}{2\pi} f_\v{k}^\pm(\omega)
    \, .
    \label{supp:eqn:simplified-matrix-gap-eqn-limiting}
\end{equation}
The correspondence between the Born--Markov description presented in the main text and the more complete Keldysh description is now clear: both approaches give rise to the same equations [i.e., Eqn.~\eqref{supp:eqn:simplified-matrix-gap-eqn}, and, in the limiting case $\abs*{V^\prime/V}$, Eqn.~\eqref{supp:eqn:simplified-matrix-gap-eqn-limiting}], but with differing expressions for the functions $F_\pm$. The most important feature of the functions $F_\pm$ in the Born--Markov approach was that they had opposite sign. In fact, we showed that when $\Gamma_1=\Gamma_2$, the functions satisfied the exact relation $F_+ = -F_-$. Evidently, for our conclusions to remain valid, the more complete description presented here must reproduce the same phenomenology in order to get superconductivity in the presence of repulsive interactions.

The phase diagram as predicted by the Keldysh description is shown in Fig.~\ref{fig:keldysh-Vc}. The two functions $F_\pm$ do indeed satisfy $F_+ \simeq -F_-$, at least in the limit of small driving ($\Omega \lesssim |\mu|$). The qualitative similarities between the two approaches are apparent. What the more complete description reveals is that this symmetry is in fact not present for all driving strengths, breaking down for sufficiently large driving powers. In addition, we also observe a breakdown of the monotonic decay of the critical coupling strength in the same large-drive regime. This is the regime in which we expect the Born--Markov results derived from~\eqref{supp:eqn:born-markov-eom} to fail: the dissipative part of the equation of motion tries to relax a generic nonequilibrium state to a thermal distribution of the electronic energy levels, while in the Keldysh description the corresponding relaxation is towards a thermal distribution of quasiparticles. The difference between the two approaches will therefore become apparent when the system quasiparticles can no longer be treated as electrons.

%%%%%%%%%%%%%%%%%%%%%%%%%%%%%%%%%%%%%%%%%%%%%%%%%%%%%%%%%%%%%%%%%%%%%%
%%%%%%%%%%%%%%%%%%%%%%%%%%%%% BROADBAND %%%%%%%%%%%%%%%%%%%%%%%%%%%%%%
%%%%%%%%%%%%%%%%%%%%%%%%%%%%%%%%%%%%%%%%%%%%%%%%%%%%%%%%%%%%%%%%%%%%%%

\section{Fast intraband relaxation}

In this section, we describe another possible regime of experimental interest: rapid intraband relaxation (with respect to the interband relaxation rate). In contrast to the case considered in the main text, the populations of the upper and lower bands may be significantly altered over a wide range of energies, allowing for smaller critical interaction strengths. In this regime, the upper and lower bands separately quasithermalise to Fermi-Dirac distributions with different chemical potentials, $\mu_2$ and $\mu_1$, which fix the number of particles in the upper and lower bands, respectively~\cite{Haug2009}.

The equations of motion in this case are~\cite{Goldstein2015}
\begin{align}
    \partial_t n_{\v{k}}^{11} &= -i\Delta_{-\v{k}} s_{-\v{k}}^{21} + i\bar{\Delta}_{-\v{k}} \bar{s}_{-\v{k}}^{21} - 2\Gamma_1 [n_{\v{k}}^{11}-n_\text{F}(\xi_{\v{k}1}, \mu_1)] \, , \label{supp:eqn:broadband-n11} \\[0.15cm]
    \partial_t n_{\v{k}}^{22} &= -i\Delta_{\v{k}} s_{\v{k}}^{21} + i\bar{\Delta}_{\v{k}} \bar{s}_{\v{k}}^{21} - 2\Gamma_2 [n_{\v{k}}^{22}-n_\text{F}(\xi_{\v{k}2}, \mu_2)] \, , \label{supp:eqn:broadband-n22} \\[0.15cm]
    \partial_t s_{\v{k}}^{21} &= (iE_\v{k}-\Gamma_{12})s_\v{k}^{21} + i\bar{\Delta}_{\v{k}}(1-n_\v{k}^{22} - n_{-\v{k}}^{11}) \, , \label{supp:eqn:broadband-s21}
\end{align}
for $\v{k} \in K_+$. The corresponding equations for $\v{k} \in K_-$ may be found by substituting $n_\text{F}(\xi_{\v{k}\alpha}, \mu_\alpha) \to n_\text{F}(\xi_{\v{k}\alpha}, \mu)$, i.e., the populations in valley $K_-$ remain thermal. If $\Delta_\v{k}=0$, i.e., in the absence of superconductivity, then the populations $n_\v{k}^{\alpha\alpha}$ will relax to the nonequilibrium distributions $n_\text{F}(\xi_{\v{k}\alpha}, \mu_\alpha)$ in the $K_+$ valley, as advertised. A relationship between the two effective chemical potentials $\mu_1$ and $\mu_2$ can be found by conserving the total number particles within the $K_+$ valley
\begin{equation}
    \sum_{\v{k} \in K_+} \left[ n_\text{F}(\xi_{\v{k}1}, \mu) + n_\text{F}(\xi_{\v{k}2}, \mu) \right] = \sum_{\v{k} \in K_+} \left[ n_\text{F}(\xi_{\v{k}1}, \mu_1) + n_\text{F}(\xi_{\v{k}2}, \mu_2) \right]
    \, .
    \label{eqn:number-conservation}
\end{equation}

The set of equations \eqref{supp:eqn:broadband-n11}--\eqref{supp:eqn:broadband-s21} can be solved to find the following steady-state population differences for the two pairing channels
\begin{align}
    1-n_{\v{k}}^{22} - n_{-\v{k}}^{11} &= \frac{1-n_\text{F}(\xi_{\v{k}2}, \mu_2) - n_\text{F}(\xi_{-\v{k}1}, \mu) }{1+ |\tilde{\Delta}_+|^2/\bar{\gamma} } \, , \\
    1-n_{-\v{k}}^{22} - n_{\v{k}}^{11} &= \frac{1-n_\text{F}(\xi_{-\v{k}2}, \mu) - n_\text{F}(\xi_{\v{k}1}, \mu_1) }{1+ |\tilde{\Delta}_-|^2/\bar{\gamma} } \, ,
\end{align}
for $\v{k} \in K_+$, where $|\tilde{\Delta}_\pm|\equiv |\Delta_\pm|^2/(E_\v{k}^2 + \Gamma_{12}^2)$, and $\bar{\gamma}^{-1} = \gamma_1^{-1} + \gamma_2^{-1}$.

Substituting into the definition of $F_+$, we arrive at
\begin{equation}
    F_+ = -\frac{E_\v{k}}{E_\v{k}^2+\Gamma_{12}^2} \frac{1}{N} \sum_{\v{k}\in K_+} \frac{1-n_\text{F}(\xi_{\v{k}2}, \mu_2) - n_\text{F}(\xi_{-\v{k}1}, \mu) }{1+ |\tilde{\Delta}_+|^2/\bar{\gamma} }
    \, .
\end{equation}
Let us define the integrated population difference as
\begin{equation}
    \mc{N}_+(\mu_2)=\frac{1}{N} \sum_{\v{k}\in K_+}\left[ 1-n_\text{F}(\xi_{\v{k}2}, \mu_2) - n_\text{F}(\xi_{-\v{k}1}, \mu) \right] < 0
    \, .
\end{equation}
This quantity is negative because the lower band in valley $K_-$ is fully occupied, $n_\text{F}(\xi_{-\v{k}1}, \mu) \simeq 1$, while the upper band in valley $K_+$ satisfies $n_\text{F}(\xi_{\v{k}2}, \mu_2) \simeq 1$ ($0$) for energies below (above) $\mu_2$.
One may interpret $P_+=2\abs*{\mc{N}_+}$ as the `polarisation' of the $K_+$ valley.
Crucially, this factor can be $O(1)$ [for $\mu_2/t=O(1)$], in contrast to the resonant driving case, which is limited to inducing population differences in the vicinity of the surface where the laser is resonant.
We therefore obtain the result
\begin{equation}
    F_+ = \frac{2\mu\mc{N}_+(\mu_2)}{4\mu^2+\Gamma_{12}^2+ |\Delta_+|^2/\bar{\gamma}}
    \, .
    \label{supp:eqn:broadband-F+}
\end{equation}
The chemical potential $\mu$ is set by an external reservoir, which also serves to drain the energy continuously being injected into our system. The quantity $\mu$ must not be confused with the effective chemical potentials $\mu_\alpha$, which determine the populations of the upper and lower bands only. The external chemical potential $\mu$ can in principle be tuned to the optimal value $2\mu = -\Gamma$, leading to a resonance. In this case,
\begin{equation}
    F_+(\Delta_+=0) = \frac{P_+}{4\Gamma_{12}}
    \, .
\end{equation}
Since the valley polarisation is (at least in principle) $O(1)$, it is the relaxation rate $\Gamma_{12}$ that sets the energy scale for $V_c$.

Considering the other pairing channel corresponding to $\Delta_-$, we find, analogous to \eqref{supp:eqn:broadband-F+},
\begin{equation}
    F_- = \frac{2\mu\mc{N}_-(\mu_1)}{4\mu^2+\Gamma_{12}^2+ |\Delta_-|^2/\bar{\gamma}}
    \, ,
\end{equation}
where we have defined the corresponding population difference
\begin{equation}
    \mc{N}_-(\mu_1)=\frac{1}{N} \sum_{\v{k}\in K_-}\left[ 1-n_\text{F}(\xi_{-\v{k}2}, \mu) - n_\text{F}(\xi_{\v{k}1}, \mu_1) \right] > 0
    \, .
\end{equation}
The polarisations $P_\pm = 2\abs*{N_\pm}$ satisfy $P_- = -P_+$ if temperature remains well below the band gap, $T \ll \delta$, giving rise to two branches for $V_c$.

To determine the magnitude of the gap in the absence of intervalley coupling ($V^\prime = 0$), we must rearrange the self-consistency condition $1=VF_+$, giving
\begin{equation}
    |\Delta_+|^2 = \frac{V}{4} \Gamma_{12} |\mc{N}_+(\mu_2)| - \frac{1}{2}\Gamma_{12}^2 = \frac{1}{4} \Gamma_{12} |\mc{N}_+(\mu_2)| (V-V_c)
    \, .
\end{equation}
The magnitude of the order parameter is plotted using this expression in Fig.~3b of the main text.

%%%%%%%%%%%%%%%%%%%%%%%%%%%%%%%%%%%%%%%%%%%%%%%%%%%%%%%%%%%%%%%%%%%%%%
%%%%%%%%%%%%%%%%%%%%%%%%%%%% BIBLIOGRAPHY %%%%%%%%%%%%%%%%%%%%%%%%%%%%
%%%%%%%%%%%%%%%%%%%%%%%%%%%%%%%%%%%%%%%%%%%%%%%%%%%%%%%%%%%%%%%%%%%%%%

\bibliographystyle{aipnum4-1}
\bibliography{references}